\documentclass[prb,aps,superscriptaddress,amsmath,amsfonts,amssymb,showpacs,floatfix,reprint,longbibliography]{revtex4-2}

\pdfoutput=1

\usepackage[subpreambles=true]{standalone}

\usepackage{mathtools}
\usepackage{newtxtext,newtxmath}
\usepackage[T1]{fontenc}
\usepackage{bm}
\usepackage{microtype}
\usepackage[version=4]{mhchem}
\usepackage{physics} 
\usepackage{graphicx}
\usepackage[percent]{overpic}
\usepackage[dvipsnames,table]{xcolor}
\usepackage{natbib}
\usepackage[utf8]{inputenc}
\usepackage{datetime}

\usepackage{multibib} 

\usepackage[linktocpage=false]{hyperref}
\usepackage[capitalise]{cleveref}

\definecolor{cblue}{RGB}{55,126,184}
\hypersetup{colorlinks=true,linkcolor=cblue,citecolor=cblue,urlcolor=cblue}

\newcommand*{\NSF}{\sigma_{\mathrm{NSF}}}
\newcommand*{\SF}{\sigma_{\mathrm{SF}}}
\newcommand*{\uvec}[1]{\bm{\hat{#1}}}
\newcommand*{\bq}{\bm{q}}
\newcommand*{\p}{\mathsf{P}} 
\newcommand{\im}{{\mathrm{i}\mkern1mu}}
\newcommand{\nsfvec}{\Omega} 
\newcommand{\eigv}{\psi}
\newcommand{\Apyro}{A}
\newcommand{\VNNSI}{V_{\mathrm{NNSI}}}
\newcommand{\VESI}{V_{\mathrm{ESI}}}

\makeatletter
\def\maketitle{
\@author@finish
\title@column\titleblock@produce
\suppressfloats[t]}
\makeatother

\begin{document}

\title{Probing Flat Band Physics in Spin Ice Systems via Polarized Neutron Scattering}

\author{K. T. K. Chung}
\affiliation{Department of Physics and Astronomy, University of Waterloo, Ontario, N2L 3G1, Canada}
\author{J. S. K. Goh}
\affiliation{Department of Physics and Astronomy, University of Waterloo, Ontario, N2L 3G1, Canada}
\affiliation{Division of Physics and Applied Physics, School of Physical and Mathematical Sciences, Nanyang Technological University, 21 Nanyang Link 637371, Singapore}
\author{A. Mukherjee}
\affiliation{Department of Physics and Astronomy, University of Waterloo, Ontario, N2L 3G1, Canada}
\author{W. Jin}
\affiliation{Department of Physics and Astronomy, University of Waterloo, Ontario, N2L 3G1, Canada}
\author{D. Lozano-G\'omez}
\affiliation{Department of Physics and Astronomy, University of Waterloo, Ontario, N2L 3G1, Canada}
\author{M. J. P. Gingras}
\affiliation{Department of Physics and Astronomy, University of Waterloo, Ontario, N2L 3G1, Canada}
\affiliation{CIFAR, MaRS Centre, West Tower 661 University Ave., Suite 505, Toronto, ON, M5G 1M1, Canada}
\date{\today}

\begin{abstract}
In this paper, we illustrate how polarized neutron scattering can be used to isolate the spin-spin correlations of modes forming flat bands in a frustrated magnetic system hosting a classical spin liquid phase. In particular, we explain why the nearest-neighbor spin ice model, whose interaction matrix has two flat bands, produces a dispersionless (i.e. ``flat'') response in the non-spin-flip (NSF) polarized neutron scattering channel, and demonstrate that NSF scattering is a highly sensitive probe of correlations induced by weak perturbations which lift the flat band degeneracy. We use this to explain the experimentally measured dispersive (i.e. non-flat) NSF channel of the dipolar spin ice compound \ce{Ho2Ti2O7}. 
\end{abstract}

\maketitle

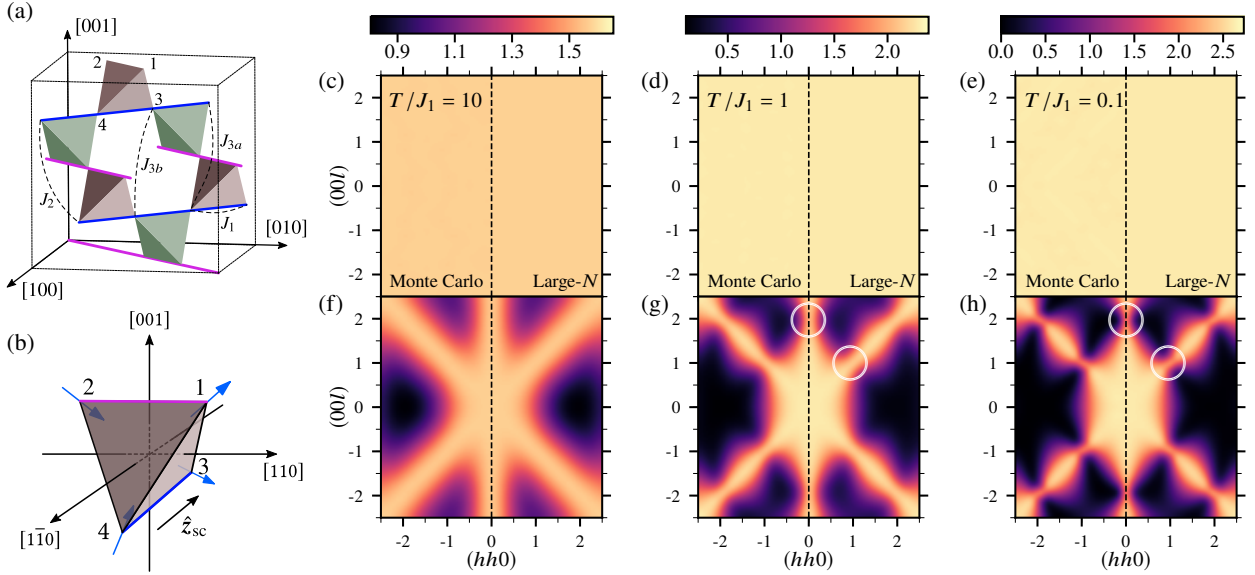
\begin{figure*}[t]
	\centering
	\begin{overpic}[width=0.46\columnwidth]{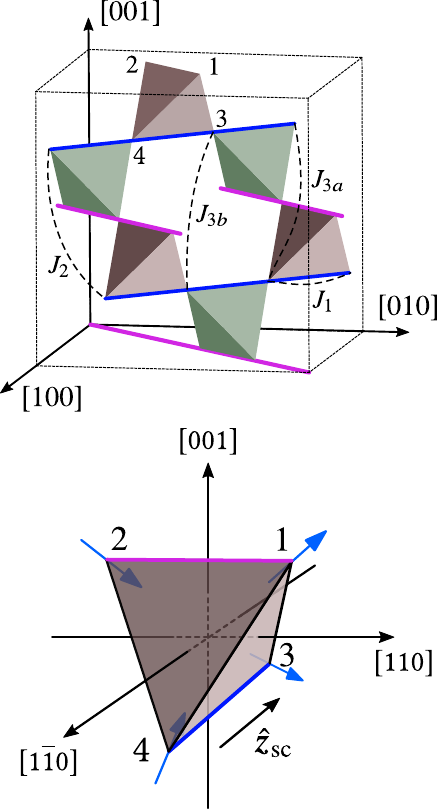}
	    \put(0,100){(a)}
	    \put(0,40){(b)}
	\end{overpic}
	\newcommand*{\figtwoLN}{\scriptsize{}\textcolor{black}{Large-$N$}}
	\newcommand*{\figtwoMC}{\scriptsize{}\textcolor{black}{Monte Carlo}}
	\begin{overpic}[width=1.46\columnwidth]{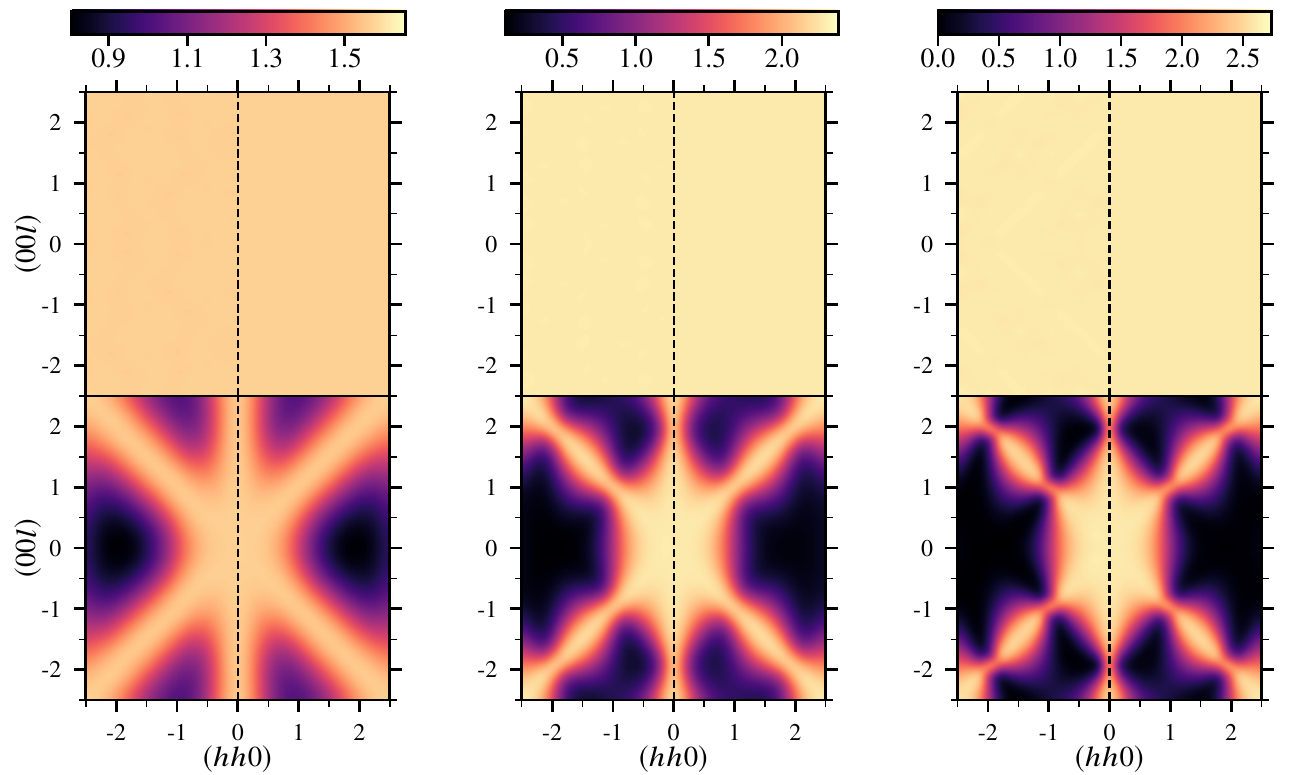}
		\put(7.5,30){\figtwoMC}
		\put(41,30){\figtwoMC}
		\put(74.5,30){\figtwoMC}
		\put(22.6,30){\figtwoLN}
		\put(56,30){\figtwoLN}
		\put(89.2,30){\figtwoLN}
		\put(7.5,49){\footnotesize $T/J_{1}=10$}
		\put(41,49){\footnotesize $T/J_{1}=1$}
		\put(74.5,49){\footnotesize $T/J_{1}=0.1$}
		\put(0,51){(c)}
		\put(34,51){(d)}
		\put(67.5,51){(e)}
		\put(0,27.5){(f)}
		\put(34,27.5){(g)}
		\put(67.5,27.5){(h)}	
		\put(89.5,22){\color{white}\circle{3.6}}
		\put(89.5,22){\color{white}\circle{3.3}}	
		\put(85.05,26.5){\color{white}\circle{3.6}}
		\put(85.05,26.5){\color{white}\circle{3.3}}	
		\put(56,22){\color{white}\circle{3.6}}
		\put(56,22){\color{white}\circle{3.3}}	
		\put(51.6,26.5){\color{white}\circle{3.6}}
		\put(51.6,26.5){\color{white}\circle{3.3}}
	\end{overpic}
	\caption{
		(a) The pyrochlore lattice with its four labeled 
			sublattices and couplings $J_{n}$ up to third neighbor. 
			$\alpha$-chains along $[\bar 1 1 0]$ discussed in the main text are highlighted in blue.
			(b) Ice-rule obeying configuration of Ising spins on a tetrahedron. (c,d,e) NSF, and (f,g,h) SF for NNSI in the $(hhl)$ plane, with pinch points indicated by white circles. 
		The left (right) half of each panel shows the Monte Carlo (large-$N$) results.
		}
	\label{fig:largeN_vs_MC}
\end{figure*}

Momentum-independent bands in electronic and magnetic systems are indicative of spatially-localized excitations of the pertinent degrees of freedom. 
Systems with such flat bands boast a huge sensitivity to  perturbations, often giving rise to exotic strongly correlated states of matter~\cite{Flatband1,Flatband2,mizoguchi2019}. Highly frustrated magnets, such as antiferromagnetically-coupled spins on kagome and pyrochlore lattices, have proven to be an inexhaustible gold mine to explore flat bands and their consequential physics, with spin ice (SI) systems~\cite{Bramwell1495,Springer-spin-ice} providing a particular fruitful setting to do so.

Momentum-resolved probes~\cite{hoppner2013,marchenko2018,Wu2015,Petit2016,Benton2016} are the most direct methods to study flat bands experimentally,  with neutron scattering being the method of choice for magnetic systems~\cite{Petit2016,Benton2016}. 
The neutron moment (spin) is sensitive to two of the three components of the local magnetic field produced by the material's magnetic moments. 
Neutron spin polarization analysis separates the moments' correlations into two channels, referred to as the spin-flip and non-spin-flip (SF and NSF, respectively)~\cite{Moon1969,Lovesey1984}. 

The flat bands of classical SI~\cite{Gingras2000,isakov2005} give rise to a low-temperature collective paramagnetic Coulomb phase~\cite{Henley2005,Henley2010,Castelnovo2012} whose emergent gauge structure is signalled in reciprocal space by `pinch points' in the neutron cross sections~\cite{Henley2005,Henley2010,Castelnovo2012}.
In SI, pinch points were first~\cite{fennell2009} experimentally investigated in \ce{Ho2Ti2O7}~\cite{harris1997,Bramwell2001,Clancy2009} using polarized neutron scattering~\cite{fennell2009}. 
In this compound, the SF channel displays pinch points in the $(hhl)$ scattering plane, reflecting singularities in the spin-spin correlations which are understood by mapping the spins to a divergence-free vector field ${\bm B}$~\cite{Henley2005,Henley2010,Castelnovo2012,Conlon2010}. 
Conversely, NSF scattering in  Ho$_2$Ti$_2$O$_7$~\cite{fennell2009} exhibits only broad diffuse features in $(hhl)$ and has received minimal attention.

Interestingly, it has been noted several times~\cite{fennell2009,flicker2011,benton2016-1,kato2015,Castelnovo2019} that nearest-neighbor spin ice (NNSI), a foundational minimal model~\cite{Anderson1956,harris1997,Bramwell1495} hosting a Coulomb phase and thus pinch points~\cite{Henley2005,Henley2010,Castelnovo2012}, displays a momentum-independent (i.e.~``flat'') NSF $(hhl)$ scattering intensity. 
Given that the NNSI model has been extensively studied and is  well-understood~\cite{isakov2004,Henley2005,Henley2010,Gingras2011,Castelnovo2012,Gingras2014,Springer-spin-ice}, it is surprising that this fact has not been scrutinized in any detail.
This raises the questions addressed in this work: how is the flat NSF intensity of NNSI related to its flat bands~\cite{Gingras2000,isakov2005}, what is the underlying physics of this relationship, and how does the NSF acquire dispersion when interactions beyond nearest-neighbor are introduced?

Using Monte Carlo simulations and a large-$N$ approximation, we confirm that NNSI and an extended spin ice (ESI) model~\cite{Rau2016,udagawa2016} exhibit a flat NSF at \emph{all} temperatures, but with a magnitude rising monotonically as temperature decreases [see Fig.~\ref{fig:largeN_vs_MC}(c,d,e)]. 
The NSF intensity in the $(hhl)$ plane is shown to directly probe fluctuations of modes constructed from the flat band eigenvectors of the interaction matrix.
We show how dispersion in the NSF, which develops when moving away from the ESI model, is a sensitive indicator of the dispersion acquired by 
the originally flat bands, and explain how a dispersive NSF arises in Ho$_2$Ti$_2$O$_7$~\cite{fennell2009,Chang2010}.


\emph{Model and methods}\,\,\,\textemdash\, 
We consider a pyrochlore lattice consisting of $L^3$ face-centered cubic (FCC) unit cells with four sites per cell [\cref{fig:largeN_vs_MC}(a,b)] and periodic boundaries (see the Supplemental Material (SM)~\cite{SM} for conventions). 
Each site $i$ of the pyrochlore lattice hosts a classical Ising spin, $s_i=\pm 1$, whose magnetic moments $\cramped{\bm{\mu}_i \propto s_i \uvec{z}_{i}}$ are constrained along the local cubic $[111]$ axes $\uvec{z}_i$. 
We consider a spin Hamiltonian with interactions between first, second, and third (class $a$, but not class $b$~\cite{Rau2016,Wills_2006,delmaestro2007}) nearest-neighbors [see \cref{fig:largeN_vs_MC}(a)],
\begin{equation}
   H
      =
    J_1\sum_{\mathclap{\langle i,j \rangle}}
    s_{i} s_{j} 
    + 
    J_2 \sum_{\mathclap{\langle\!\!\langle i,j \rangle\!\!\rangle}}
    s_{i} s_{j}
    + 
    J_{3a} \sum_{\mathclap{\langle\!\!\langle\!\!\langle i,j \rangle\!\!\rangle\!\!\rangle_{a}}}
    s_{i} s_{j}
    \,,
    \label{eq:Hamiltonian}
\end{equation}
with $J_1 > 0$.
Restricting to the line of parameter space $J_2 = J_{3a} \equiv J'$, one obtains the ESI model~\cite{Rau2016,udagawa2016}, for which an extensive number of spin configurations obeying the two-in/two-out `ice rules' [illustrated in \cref{fig:largeN_vs_MC}(b)] are energetically degenerate~\cite{Bramwell1495,Henley2005,Henley2010,Castelnovo2012,harris1997,Ramirez1999}. For $-0.5 < J'/J_1 < 0.25$ (including NNSI at $J'=0$), these configurations are the ground states~\cite{Rau2016,udagawa2016}, and we refer to this restricted range as ``the ESI line''.

In polarized neutron scattering (sc) experiments with incident neutron polarization axis $\uvec{z}_{\mathrm{sc}}$, one defines an orthonormal basis for each scattering wavevector $\cramped{\bq\perp\uvec{z}_{\mathrm{sc}}}$, with $\cramped{\uvec{x}_{\mathrm{sc}} \equiv \hat{\bq}}$ and $\cramped{\uvec{y}_{\mathrm{sc}} \equiv \uvec{z}_{\mathrm{sc}} \times \uvec{x}_{\mathrm{sc}}}$~\cite{SM}. 
The scattered neutron moment is only sensitive to the $\uvec{y}_{\mathrm{sc}}$ and $\uvec{z}_{\mathrm{sc}}$ components of the $\bm{\mu}_i$, whose correlations are separated by filtering the scattered beam by neutron spin polarization~\cite{Moon1969,Lovesey1984}. This gives energy-integrated SF and NSF cross sections~\cite{fennell2009,Chang2010,chang2012}, respectively proportional to the following two structure factors~\cite{SM}:
\begin{align}
\sigma_{\mathrm{SF}}(\bq) 
&= \sum_{\mu, \nu} 
(\uvec{z}_{\mu} \cdot \uvec{y}_{\mathrm{sc}}) \,
\langle s_{\mu}^* (\bq) \, s_{\nu} (\bq)\rangle  (\uvec{z}_{\nu} \cdot \uvec{y}_{\mathrm{sc}}) \label{eq:sf_cs}  \,, 
\\
\NSF(\bq)  &= \sum_{\mu, \nu} 
(\uvec{z}_{\mu} \cdot \uvec{z}_{\mathrm{sc}}) \,
\langle s_{\mu}^* (\bq) \, s_{\nu} (\bq)\rangle (\uvec{z}_{\nu} \cdot \uvec{z}_{\mathrm{sc}}) \label{eq:nsf_cs} \, .
\end{align}
Here,  $\mu,\nu$ label the four 
FCC sublattices [\cref{fig:largeN_vs_MC}(a,b)], and $s_\mu(\bq)\equiv\frac{1}{\sqrt{L^3}}\sum_{i\in \mu}s_i\, e^{-\mathrm{i}\,\bq\cdot\bm{r}_i}$ are the Fourier-transformed Ising variables (see~\cite{SM} for conventions).
Our focus is the experimentally preferred $(hhl)$ plane~\cite{fennell2009,Chang2010,chang2012,Petit2016} with $\uvec{z}_{\mathrm{sc}}\!\equiv\![\overline{1}10]$ [see Fig.~\ref{fig:largeN_vs_MC}(b)].

To calculate the spin-spin correlations in \cref{eq:sf_cs,eq:nsf_cs}, we employ the \mbox{large-$N$} approximation \cite{Garanin1996,Conlon2010} (see~\cite{SM} for details), previously successfully used to expose many key aspects of SI physics~\cite{Canals2001,isakov2004,mizoguchi2018,lantagne2018}. We write \cref{eq:Hamiltonian} as ${H = \frac{1}{2}\sum_{ij}s_i V_{ij} s_j}$, where
$V$ is the interaction matrix, with $V_{ij}$ the coupling between sites $i$ and $j$, and $V_{ii}\equiv \varepsilon$ chosen to set the minimum eigenvalue of $V$ to zero~\cite{Conlon2010,SM}. 
The large-$N$ correlation matrix, $\mathcal{G}_{ij} \equiv \langle{s_i s_j}\rangle = [\lambda \openone + \beta V]^{-1}_{ij}$, is $4\times 4$ block diagonal in $\bq$-space~\cite{isakov2004},
\begin{equation}
    \mathcal{G}_{\mu\nu}(\bq) \equiv \langle{s_\mu^*(\bq) s_\nu(\bq)}\rangle = [\lambda \openone_{4\times 4} + \beta V(\bq)]^{-1}_{\mu\nu} \, .
    \label{eq:G-large-N}
\end{equation}
Here, $\beta=1/T$ with $T$ the temperature ($k_\mathrm{B}\equiv 1$), and $\lambda$~is a positive temperature-dependent Lagrange multiplier determined self-consistently~\cite{Conlon2010,isakov2004,lantagne2018} by the 
saddle-point condition $\Tr\mathcal{G} \equiv \sum_i \langle s_i^2 \rangle = 4L^3$ (the number of spins).


\emph{Results}\,\,\textemdash\,
Starting with NNSI $\cramped{(J'=0)}$, 
$\SF$ [\cref{fig:largeN_vs_MC}(f,g,h)] displays a distinct 
scattering pattern in $(hhl)$, with pinch points (white circles)
developing for $\cramped{T/J_1 \lesssim 1}$ signaling the onset of the Coulomb phase.
In contrast, $\NSF$ [\cref{fig:largeN_vs_MC}(c,d,e)] is  ${\bq}$-independent at all temperatures,  with intensity rising monotonically as temperature decreases. Analogous results are obtained for models on the ESI line \cite{SM,footnote_FM}.  In all cases, a flat $\NSF$ is only observed for $\cramped{\bq \in (hhl)}$ and symmetry-equivalent planes --- cf.~the non-flat NNSI $\NSF$ for $\cramped{\bq\in(h0l)}$ in~\cite{SM}.

To investigate the origin of this flatness, let $\bm{\nsfvec}$ be a 4-component vector with components $\cramped{\nsfvec_\mu\equiv(\uvec{z}_{\mathrm{sc}}\cdot\uvec{z}_\mu)}$ in the sublattice basis, and normalized components denoted  $\cramped{\hat{\nsfvec}_\mu\equiv \nsfvec_\mu/\abs{\bm{\nsfvec}}}$,
with which we rewrite \cref{eq:nsf_cs} as
\begin{equation}
    \NSF(\bq) = \abs{\bm{\nsfvec}}^2\, \langle \vert \hat{\nsfvec}_\mu s_\mu(\bq) \vert^2 \rangle\,,
    \label{eq:NSF-p-s}
\end{equation}
with implied summation over repeated index $\mu$. We refer to the normalized linear combination of spin variables $\hat{\nsfvec}_\mu s_\mu(\bq)$ as a \emph{mode} (one mode for each $\bq$), and interpret $\langle \vert \hat{\nsfvec}_\mu s_\mu(\bq) \vert^2 \rangle$ as its thermal occupation value (TOV). Crucially, when $\uvec{z}_{\mathrm{sc}}\parallel [\bar{1}10]$, the Ising moments on sublattices 1 and 2 lie orthogonal to $\uvec{z}_{\mathrm{sc}}$ [see \cref{fig:largeN_vs_MC}(b)] so that 
$\bm{\nsfvec} = \sqrt{2/3}\,(0,0,1,-1)$ and $\abs{\bm{\nsfvec}}^2=4/3$ --- 
only spins on sublattices 3 and 4 contribute to NSF scattering in the $(hhl)$ plane. 

To evaluate $\NSF$ in \cref{eq:NSF-p-s}, we begin with a spectral decomposition of $V$~\cite{Gingras2000,reimers1991}, $\cramped{V_{ij} = \sum_{\bq,n} \epsilon_n(\bq) [\hat{\psi}_n(\bq)]_i [\hat{\psi}_n(\bq)]^*_j}$. The normalized eigenvectors 
$\hat{\bm{\eigv}}_n(\bq)$ ($\cramped{n=1,2,3,4}$) define the \emph{normal modes} $\cramped{\tilde{s}_n(\bq) \equiv \sum_i[\hat{\eigv}_n(\bq)]_i s_i}$. The corresponding eigenvalues $\cramped{\epsilon_n(\bq)\geq 0}$ are the normal mode energies, forming four bands indexed by $n$, and the Hamiltonian is $\cramped{H = \frac{1}{2}\sum_{\bq,n}
\epsilon_n(\bq) \vert \tilde{s}_n(\bq) \vert^2}$. 
The correlation matrix $\mathcal{G}$ [\cref{eq:G-large-N}] 
shares the eigenvectors of $V$ and its eigenvalues are precisely the normal mode TOVs,
$\cramped{\langle \vert\tilde{s}_n(\bq)\vert^2\rangle = [\lambda+\beta \epsilon_n(\bq)]^{-1}}$. 

The interaction matrix for NNSI is  $\cramped{\VNNSI\equiv J_1(\Apyro+2 \openone)}$ 
and, for ESI, $\cramped{\VESI\equiv \VNNSI + J'(\Apyro^2 - 2\Apyro-8\openone)}$~\cite{mizoguchi2019}. Here, $\Apyro$ is the $4L^3\times 4L^3$ pyrochlore nearest-neighbor adjacency matrix encoding the connectivity of the lattice~\cite{mizoguchi2019}, which is block diagonal in $\bq$-space, with each $4\times 4$ block denoted $\Apyro(\bq)$. Importantly, $\Apyro$ has two flat bands at the bottom of its spectrum due to the geometric frustration of this lattice. 
Since $\VNNSI$ and $\VESI$ are polynomials of $\Apyro$, they share its eigenvectors and inherit zero-energy flat bands~\cite{mizoguchi2019} and positive-energy dispersive bands.
Importantly, $\hat{\bm{\nsfvec}}$ in \cref{eq:NSF-p-s} is a flat band eigenvector of $\Apyro(\bq)$ for all $\bq \in(hhl)$~\cite{SM}, and thus also of $\VNNSI(\bq)$ and $\VESI(\bq)$.
From this, it follows that the aforementioned modes $\{\hat{\nsfvec}_\mu s_\mu(\bq)\}$ probed by NSF scattering are \emph{flat band normal modes} of $\VNNSI$ and $\VESI$ with $\cramped{\epsilon_n(\bq)=0}$,  whose TOV are $\langle \vert \hat{\nsfvec}_\mu s_\mu(\bq) \vert^2 \rangle = \lambda^{-1}$.
Therefore, \cref{eq:NSF-p-s} yields 
\begin{equation}
    \sigma_{\mathrm{NSF}}(\bq) = \frac{4}{3\lambda}\,.
    \label{eq:43lambda}
\end{equation}
Thus the $(hhl)$ NSF is \mbox{$\bq$-independent}, increasing monotonically from $4/3$ in the high-$T$ paramagnetic phase (TOV of all modes equal to 1 $\Rightarrow \lambda =1$) to $8/3$ in the low-$T$ Coulomb phase (TOV of dispersive band modes equal to 0, TOV of flat band modes equal to 2 $\Rightarrow \lambda=1/2$)~\cite{SM} as seen in \cref{fig:largeN_vs_MC}(c-e).


\emph{Coulomb phase interpretation}%
\,
\textemdash
\,
We established above that, irrespective of the $V_{ij}$ considered, $\NSF$ probes for each $\cramped{\bq\in(hhl)}$ a mode $\cramped{\hat{\nsfvec}_\mu s_\mu(\bq)}$ constructed from the components of the flat band eigenvector $\hat{\bm{\nsfvec}}$ of $\Apyro(\bq)$.
The modes $\cramped{\{\hat{\nsfvec}_\mu s_\mu(\bq)\}}$ are energetically degenerate for $\VNNSI$ and $\VESI$, resulting in a flat NSF.
Given that the physics of NNSI and ESI is controlled entirely by the spectral properties of $\Apyro$, it will prove useful to adopt a terminology differentiating between modes constructed from the dispersive band eigenvectors of $\Apyro$ and those constructed from its flat band eigenvectors. 
To set up this terminology, we first focus on the long-wavelength limit describing the coarse-grained Coulomb phase physics of $\VNNSI$ and $\VESI$. In this limit, the pertinent normal modes are obtained by an orthonormal change of basis~\cite{Henley2005,Conlon2010},
\begin{align}
    Q(\bq) \equiv \frac{1}{2} \sum_{\mu} s_\mu(\bq)
    \,, \quad
    \bm{B}(\bq) \equiv \sqrt{\frac{3}{4}} \sum_\mu s_\mu(\bq)\uvec{z}_\mu\, .
        \label{eq:QB}
\end{align}
In direct-space, $Q$ and $\bm{B}$ are respectively akin to a charge and a 3-component vector field.
The long-wavelength dispersive band modes are $Q(\bq)$ and $\hat{\bq}\cdot\bm{B}(\bq)$, which are thermally depopulated at low temperature~\cite{Conlon2010}. 
The long-wavelength flat band modes are the two remaining components of $\bm{B}(\bq)$, which lie in the plane orthogonal to $\hat{\bq}$ (spanned by $\uvec{y}_{\mathrm{sc}}$ and $\uvec{z}_{\mathrm{sc}}$). In a gauge theory like electromagnetism, the pair \{$Q(\bq)$, $\,\cramped{\hat{\bq}\cdot\bm{B}(\bq)}$\} would commonly be referred to as \emph{longitudinal} modes and the pair \{$\cramped{\uvec{y}_{\mathrm{sc}}\cdot\bm{B}(\bq)}$,   $\,\cramped{\uvec{z}_{\mathrm{sc}}\cdot\bm{B}(\bq)}$\} as \emph{transverse} modes.

We now extend this terminology to arbitrary $\bq$, where the longitudinal modes refer to modes constructed from dispersive band eigenvectors of $\Apyro$, while the transverse modes are those constructed from flat band eigenvectors of $\Apyro$. 
Expressed in this basis, $\SF$ and $\NSF$ take simple forms,
    \begin{align}
        \SF= \frac{4}{3}\langle\abs{\bm{B}(\bq)\cdot\uvec{y}_{\mathrm{sc}}}^2\rangle
        \,,\quad
        \NSF=\frac{4}{3}\langle\abs{\bm{B}(\bq)\cdot\uvec{z}_{\mathrm{sc}}}^2\rangle.
        \label{eq:sf-nsf-B}
    \end{align} 
The modes whose TOV appear in \cref{eq:sf-nsf-B} are both transverse modes \emph{at long wavelength}, which is reflected in the equal intensity of $\SF$ and $\NSF$ seen in \cref{fig:largeN_vs_MC} for small $\bq$. 
However, for \emph{larger} wavevectors the SF intensity drops to zero at low temperature, indicating that $\cramped{\bm{B}(\bq)\cdot\uvec{y}_{\mathrm{sc}}}$ is now a \emph{longitudinal} mode. 
The NSF is flat throughout $(hhl)$ because $\cramped{\bm{B}(\bq)\cdot\uvec{z}_{\mathrm{sc}}\equiv \hat{\Omega}_\mu s_\mu(\bq)}$ is a transverse mode for \emph{all} $\bq$ in this plane, since $\hat{\bm{\Omega}}$ is a flat band eigenvector of $\Apyro(\bq)$. 
The lesson is that while the SF probes both transverse and longitudinal modes, resulting in pinch points, the $(hhl)$ NSF probes \emph{solely} transverse modes. 
In the aforementioned long-wavelength theory of 
 NNSI~\cite{Henley2005,Henley2010,Conlon2010} built from $Q$ and ${\bm B}$, the physics is controlled by $\lambda$ and a screening length $\xi$~\cite{Henley2010} (alternatively, the charge density), which controls the width of the pinch points~\cite{Henley2010}. 
Since the NSF is blind to the longitudinal modes --- $\xi$ does not appear in \cref{eq:43lambda} --- a calibrated measurement of the temperature dependence of $\NSF$ could afford a direct experimental determination of $\lambda(T)$ that characterizes the long-wavelength theory.

\emph{Chain Correlations}\,\,\textemdash\,\,
We now turn to the direct space interpretation of the modes $\cramped{\{\hat{\nsfvec}_\mu s_\mu(\bq)\}}$ probed by the NSF. 
First, note that sublattices 3 and 4 lie on $L^2$  `$\alpha$-chains' running along $[\bar{1}10]$ ~\cite{Ruff2005,Hiroi2003,Clancy2009} [\cref{fig:largeN_vs_MC}(a) blue lines], which form a 2D isosceles triangular lattice perpendicular to $[\bar{1}10]$~\cite{Ruff2005,Hiroi2003,Higashinaka2005,mcclarty2015}. 
We define an $\alpha$-chain's polarization $\cramped{\p_\alpha\equiv\tfrac{1}{\sqrt{L}}\left( \sum_{i\in\alpha}s_i\,\uvec{z}_i\right)\cdot\uvec{z}_\mathrm{sc}}$ and, for each $\bq \in (hhl)$, its Fourier transform $\cramped{\p(\bq)\equiv\frac{1}{\sqrt{L^2}}\sum_{\alpha}\p_\alpha\,e^{-\mathrm{i}\,\bq\cdot\bm{R}_\alpha}}$, where $\bm{R}_\alpha$ are the $\alpha$-chain coordinates in the $[hhl]$ plane. 
One easily obtains~\cite{SM} 
\begin{align}
    \cramped{
    \sigma_{\mathrm{NSF}}(\bq)
    = \langle \abs{\p(\bq)}^2\rangle 
    = \frac{1}{L^2}\sum_{\alpha,\alpha'}
    \langle \p_\alpha \p_{\alpha'} \rangle\,
    e^{-\mathrm{i}\bq\cdot(\bm{R}_{\alpha'}-\bm{R}_{\alpha})} \, ,
    }
    \label{eq:alpha}
\end{align}
i.e. $\NSF(\bq)$ is the Fourier transform of the chain-chain correlation function $\langle \p_\alpha \p_{\alpha'}\rangle$. 
Comparing \eqref{eq:alpha} with \cref{eq:NSF-p-s,eq:sf-nsf-B}, we see that the transverse mode $\hat{\nsfvec}_\mu s_\mu(\bq)$ probed by $\NSF(\bq)$ is the Fourier transformed $\alpha$-chain polarization,  $\hat{\nsfvec}_\mu s_\mu(\bq) = \sqrt{3/4}\,\p(\bq)$. 
A geometrical interpretation of the $(hhl)$ NSF follows: trivially, sublattices 1 and 2 do not contribute because their Ising moments lie perpendicular to the $\alpha$-chains along which the neutrons are polarized, thus the NSF isolates the spin correlations of sublattices 3 and 4. 
These `3-4' chains are the support of flat band eigenvectors of $\Apyro$ whose direct-space components alternate sign along a chain and are zero on all other sites~\cite{bergman2008}. 
This is the geometrical origin of why $\cramped{\hat{\nsfvec}\propto(0,0,1,-1)}$
is a flat band eigenvector of $\Apyro(\bq)$ for all $\bq\in(hhl)$~\cite{SM}. 
The $\bq$-independent NSF on the ESI line indicates that 
$\langle \p_\alpha \p_{\alpha'}\rangle = (4/3\lambda)\delta_{\alpha\alpha'}$ 
--- the 
$\alpha$-chains are uncorrelated from each other at all temperatures, and 
the uniform intensity reflects \emph{intra-chain} $\langle \p_\alpha^2\rangle = (4/3\lambda)$. 
It would be interesting to investigate how this last result arises order-by-order in an approximation-free direct-space high-temperature expansion~\cite{Harris1992} of the original SI and ESI models.


\emph{$\NSF$ off the ESI line}\,\,\textemdash\,\, 
Contrasting with the previous discussion, a dispersive $\NSF$ implies non-trivial \emph{inter-chain} correlations, which we now consider.
Together, Eqs.~\eqref{eq:nsf_cs}, \eqref{eq:G-large-N}, and \eqref{eq:NSF-p-s} give
\begin{equation}
    \NSF(\bq) = \abs{\bm{\nsfvec}}^2 \,\, \hat{\nsfvec}_\mu \mathcal{G}_{\mu\nu}(\bq)\hat{\nsfvec}_\nu\,,
    \label{eq:NSF_p-G-p}
\end{equation}
which yields $\NSF=\tfrac{4}{3}[\mathcal{G}_{33}(\bq)-\mathcal{G}_{34}(\bq)]$. In Fig.~\ref{fig:linecuts}, we show line cuts (solid lines) of $\mathcal{G}_{33}(\bq)$, $\mathcal{G}_{34}(\bq)$, and $\NSF(\bq)$ along $(hh2)$ for four $(J_2,J_{3a})$ parameter choices at $T/J_1 = 0.1$. 
The NNSI case is shown in (a), while (b) corresponds to a point on the $J_2=J_{3a}$ ESI line, both of which have flat bands and thus exhibit a flat $\NSF$. 
Conversely, (c) and (d) illustrate that for slight perturbations off the ESI line weakly lifting the flat band degeneracy, $\NSF$ departs significantly from flatness, indicating the development of inter-chain correlations $\langle \p_\alpha \p_{\alpha'}\rangle$. 
This demonstrates that the NSF provides
a sensitive probe of perturbations that lift the original flat band degeneracy and make the transverse modes dispersive.

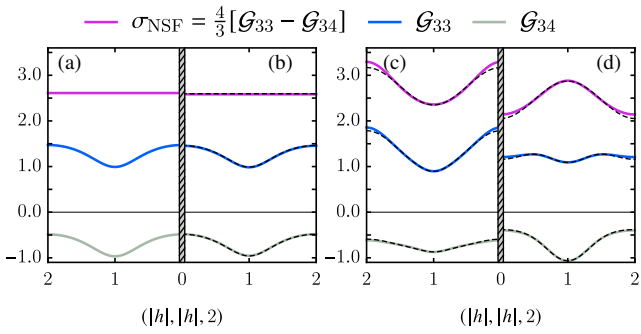
\begin{figure}[!t]
    \centering
    \begin{overpic}[width=0.98\columnwidth]{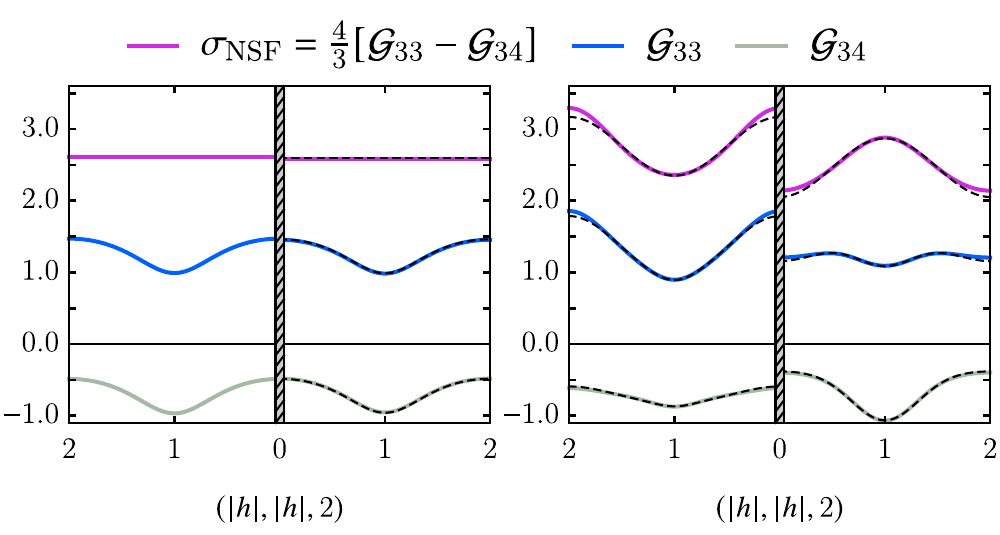}
    \put(08.5,41.5){\footnotesize (a)}
    \put(41.5,41.5){\footnotesize (b)}
    \put(59.5,41.5){\footnotesize (c)}
    \put(93.0,41.5){\footnotesize (d)}
    \end{overpic}
    \caption{Line cuts along $(hh2)$ of the $\mathcal{G}_{33}(\bq)$ and     $\mathcal{G}_{34}(\bq)$ sublattice correlations, as well as $\NSF(\bq)$, for $T/J_1 = 0.1$, with $(J_2/J_1,J_{3a}/J_1) = $ (a) $(0,0)$, (b) $(0.1,0.1)$, (c) $(0.001,-0.001)$, and (d) $(-0.001,0.001)$. Solid lines show the numerically exact values from \cref{eq:G-large-N} while dashed lines show the first-order perturbative calculation. 
       }
    \label{fig:linecuts}
\end{figure}

To expose how $\NSF$ develops dispersion, we consider perturbations $V_p$ away 
from NNSI, $\cramped{V = \VNNSI + V_p}$, where $V_p$ is a linear combination of interaction matrices with energy scale much smaller than $J_1$.  
From \cref{eq:G-large-N}, expanding  $\cramped{\lambda \equiv \lambda_0 + \lambda_p}$, $\mathcal{G}$ satisfies $\cramped{\mathcal{G}^{-1} = \mathcal{G}_0^{-1} + {\Sigma}}$, where $\cramped{\mathcal{G}_0\equiv[\lambda_0\openone + \beta \VNNSI]^{-1}}$ is the unperturbed correlation matrix and $\cramped{\Sigma \equiv [\lambda_p \openone + \beta V_p]}$ contains the perturbing terms, yielding an expansion $\cramped{\mathcal{G} = \mathcal{G}_0 - \mathcal{G}_0 \Sigma \mathcal{G}_0 + \cdots}$ (see~\cite{SM} for details). 

At low temperature, $\mathcal{G}_0$ is proportional to the projector onto the flat bands of $\VNNSI$ \cite{Henley2005}, making $\mathcal{G}_0\Sigma \mathcal{G}_0$ the projection of $\Sigma$ into the transverse mode subspace. Since $\hat{\bm{\nsfvec}}$ is an eigenvector of $\mathcal{G}_0(\bq)$ with eigenvalue $\lambda_0^{-1}$, \cref{eq:NSF_p-G-p} yields to first order
\begin{equation}
    \NSF(\bq) \approx \frac{4}{3\lambda_0} \left(1 - \frac{1}{\lambda_0}\hat{\nsfvec}_\mu\Sigma_{\mu\nu}(\bq)\hat{\nsfvec}_\nu\right) .
    \label{eq:NSF_perturb}
\end{equation}
The first order correction yields a dispersive contribution to the NSF, reflecting how $V_p$ causes the transverse modes $\{\hat{\Omega}_\mu s_\mu(\bq)\}$ to become dispersive. \cref{fig:linecuts}(b,c,d) compares \cref{eq:NSF_perturb} (dashed lines) with the exact calculation (solid lines), demonstrating that \cref{eq:NSF_perturb} accurately captures the departure from flatness when perturbing off the ESI line.

\begin{figure}[!t]
    \centering
    \includegraphics[width=0.97\columnwidth]{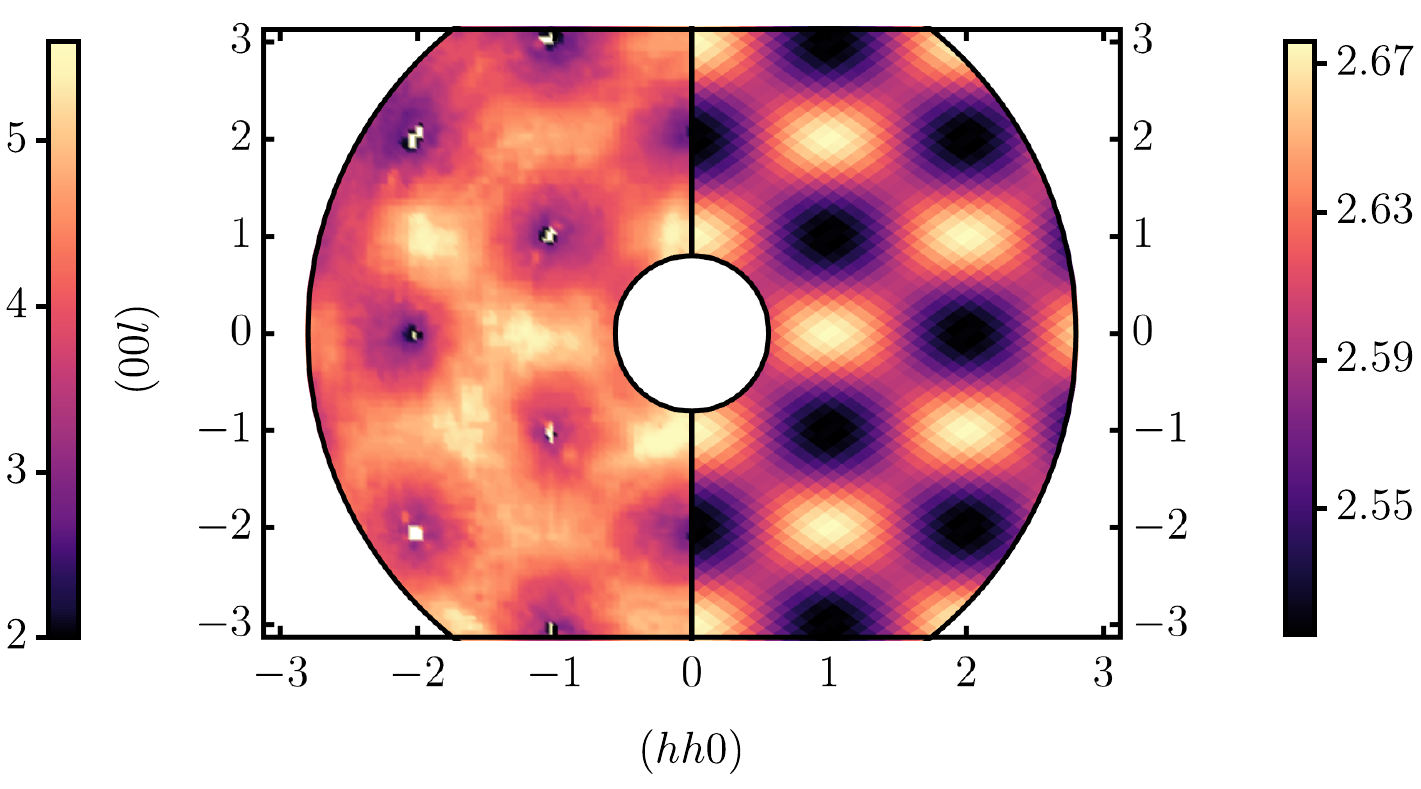}
    \caption{Left: The NSF channel intensity (in arbitrary units) measured via polarized neutron scattering in \ce{Ho2Ti2O7} (Figure adapted from Ref.~\cite{fennell2009}). Right: $\NSF$ calculated via \cref{eq:NSF_DSI_perturb}
    taking $J_{\mathrm{dip}}/J_1 = 0.03$ and $T/J_1 = 0.1$.}
    \label{fig:NSF_DSI_perturb}
\end{figure}


\emph{NSF of dipolar spin ice}\,\,\textemdash\,\, 
A natural physically relevant setting to explore how $\NSF$ acquires dispersion is to consider a dipolar spin ice (DSI)  model in which the long-range dipolar interactions added to the NNSI model are treated perturbatively. 
This is pertinent to the weak-moment non-Kramers Pr$^{3+}$-based [Pr$_2$(Sn,Hf,Zr)$_2$O$_7$] quantum SI  materials~\cite{Onoda2011,Gingras2014,Rau2019}, and is also relevant to Ho$_2$Ti$_2$O$_7$~\cite{harris1997,Bramwell2001,Clancy2009} and  Dy$_2$Ti$_2$O$_7$~\cite{henelius2016,Borzi2016} DSI compounds. 
We take $V_p=J_{\mathrm{dip}}\mathcal{D}$, where $\mathcal{D}$ is the dimensionless form of the dipolar interaction matrix~\cite{SM}, which we decompose using the `projective equivalence' of Ref.~\cite{isakov2005}, $\mathcal{D}=c\mathcal{P}+\Delta$. 
Here, $c$ is a proportionality constant and $\mathcal{P}$ is the projector into the \emph{dispersive} band eigenspace of 
$\Apyro$ --- the longitudinal mode subspace --- 
containing the long-range $1/R^3$ portion of the interaction. $\Delta$ contains small short-range corrections decaying as  $1/R^5$~\cite{isakov2005,SM}. 
The key observation is that $\mathcal{P}_{\mu\nu}(\bq)\hat{\nsfvec}_\nu = 0$ because $\hat{\bm{\nsfvec}}$ is a flat band eigenvector for $\bq \in (hhl)$. 
Therefore, the leading-order dispersive correction to the originally flat $\NSF$ in \cref{eq:NSF_perturb} is {\emph{entirely}} due to the weak short-range corrections contained in $\Delta$, giving
\begin{equation}
    \!\!\cramped{\NSF(\bq)\!\approx\!\frac{4}{3\lambda_0} \left(1 - \frac{1}{\lambda_0}\hat{\nsfvec}_\mu [\lambda_p \delta_{\mu\nu} + \beta J_{\mathrm{dip}}\Delta_{\mu\nu}(\bq)]\hat{\nsfvec}_\nu\right)\!.}
    \label{eq:NSF_DSI_perturb}
\end{equation}
We therefore conclude that the $(hhl)$ NSF of this weak-moment DSI model is \emph{insensitive} to the long-range portion of the interaction contained in $\mathcal{P}$, which encodes the $1/R$ Coulomb interaction of monopoles~\cite{Castelnovo2008}. 
It is, \emph{instead}, sensitive to the dispersion of the transverse modes induced by the weak lifting of the flat band degeneracy via the interactions contained in $\Delta$~\cite{isakov2005}.
To illustrate this, we show in  \cref{fig:NSF_DSI_perturb} the experimental NSF of $\ce{Ho2Ti2O7}$~\cite{fennell2009} compared to $\NSF(\bq)$ computed using \cref{eq:NSF_DSI_perturb} with $\beta J_{\mathrm{dip}} = 0.3$ (see SM~\cite{SM} for details). The qualitative agreement between the perturbative and experimental results is reasonable because the dispersion in the flat bands of DSI is very small relative to the dipolar interaction scale $J_{\mathrm{dip}}$ --- the phenomenon of ``self-screening''~\cite{Gingras2000,isakov2005}.


\emph{Conclusion} \,\textemdash\,
In this work, we have demonstrated that for magnetic systems on the pyrochlore lattice with localized Ising moments, the NSF neutron scattering cross section in the $(hhl)$ plane directly probes fluctuations of transverse modes. 
This explains the origin of the long-noted momentum-independent NSF of classical extended spin ice (ESI) systems, in particular nearest-neighbor spin ice (NNSI)~\cite{fennell2009,benton2016-1,kato2015,Castelnovo2019,flicker2011}.
Furthermore, we showed that the NSF channel serves as a sensitive probe of perturbations that lift the flat band degeneracy. 
Our work illustrates that an analysis of $\NSF$ could prove a fruitful approach to parameterize interactions beyond nearest-neighbor~\cite{henelius2016,Borzi2016}, quantum fluctuations~\cite{kato2015}, and lattice strain~\cite{Pili2021} in spin ice like systems. 
Crucially, our results only rely on the moments being Ising, resulting in the isolation of the $\alpha$-chains and thus the NSF probing solely transverse modes, implying a broad range of applicability in geometrically frustrated Ising magnets.

\emph{Acknowledgements} \,\textemdash\,
We thank  Cyrus  Cerkauskas, Tom Fennell and  Jeffrey  Rau for useful and stimulating discussions. This work was supported by the NSERC of Canada and the Canada Research Chair program (M.J.P.G., Tier 1). J.G.S.K. thanks Nanyang Technological University for financial support through the CN Yang Scholars Programme. 

\let\oldaddcontentsline\addcontentsline
\renewcommand{\addcontentsline}[3]{}
%

\let\addcontentsline\oldaddcontentsline

\clearpage

\title{Supplemental Material for ``Probing Flat Band Physics in Spin Ice Systems via Polarized Neutron Scattering'' }
\maketitle

\setcounter{page}{1} 
\setlength\topmargin{-64pt}\setlength\textheight{700pt} 
\onecolumngrid 
\numberwithin{equation}{section}
\numberwithin{figure}{section}

\tableofcontents

\newpage 

\section{Conventions}
\label{SM-Conventions}

\subsection{Sublattices}
\label{SM-Sublattices}

We use a face centered cubic (FCC) primitive cell with four atoms per unit cell to describe the pyrochlore lattice structure. With respect to a Cartesian coordinate system, denoting a vector $a \uvec{x} + b \uvec{y}+ c \uvec{z} \equiv [a,b,c]$, we use the following convention for the primitive displacement vectors:
\begin{equation*}
    \bm{a}_1 = a_0[1,1,0]/2\,, \quad 
    \bm{a}_2 = a_0[1,0,1]/2\,, \quad 
    \bm{a}_3 = a_0[0,1,1]/2\,,
\end{equation*}
where $a_0$ is the edge length of the conventional cubic unit cell whose basis vectors are $a_0[1,0,0]$, $a_0[0,1,0]$, and $a_0[0,0,1]$, containing 16 pyrochlore sites. We take the four sublattice positions within the FCC unit cell to be given by 
\begin{equation*}
    \bm{c}_1 = \bm{0}\,, \quad
    \bm{c}_2 = \bm{a}_1/2\,, \quad
    \bm{c}_3 = \bm{a}_2/2\,, \quad
    \bm{c}_4 = \bm{a}_3/2\,.
\end{equation*}
The Ising moments located on the pyrochlore lattice sites are constrained to point along the local $\langle 111 \rangle$ axes $\uvec{z}_{\mu}$, where $\mu=1,2,3,4$ labels the four sublattices, [see Fig.~1(b) in the main text], given by: 
\begin{equation}
\uvec{z}_{1} = \frac{1}{\sqrt{3}}[ 1, 1, 1]\,, \quad
\hat{\bm{z}}_{2} = \frac{1}{\sqrt{3}}[-1,-1, 1]\,, \quad
\hat{\bm{z}}_{3} = \frac{1}{\sqrt{3}}[-1, 1,-1]\,, \quad
\hat{\bm{z}}_{4} = \frac{1}{\sqrt{3}}[ 1,-1,-1]\,.
\label{eq:local_z_axes}
\end{equation}

\subsection{Fourier Transforms}
\label{SM-Fourier-Transforms}

The Ising spin variables can be indexed by the unit cell position $\bm{r}$ and sublattice index $\mu$, which we denote $s_\mu(\bm{r})$. We always assume a periodic lattice with $L^3$ FCC unit cells. Our Fourier transform convention is 
\begin{equation}
    s_\mu(\bq) = \frac{1}{\sqrt{L^3}}\sum_{\bm{r}} s_\mu(\bm{r}) e^{-\im\bq\cdot(\bm{r}+\,\bm{c}_\mu)},
    \label{eq:SM-s-FT}
\end{equation}
For quantities carrying two position indices, e.g. $A_{\mu\nu}(\bm{r},\bm{r}')$, the Fourier transforms are given by
\begin{equation*}
    A_{\mu\nu}(\bm{q},\bm{q}') = \frac{1}{L^3} \sum_{\bm{r},\bm{r}'} A_{\mu\nu}(\bm{r},\bm{r}') e^{-\im\bq\cdot(\bm{r}'+\,\bm{c}_\nu)} e^{\im\bq\cdot(\bm{r}+\,\bm{c}_\mu)}.
\end{equation*}
For translationally invariant quantities such as interaction matrices and correlation matrices,  $A_{\mu\nu}(\bm{r},\bm{r}') = A_{\mu\nu}(\bm{0},\bm{r}'-\bm{r})\equiv A_{\mu\nu}(\bm{0},\bm{R})$, and the matrices are $4\times 4$ block diagonal in the Fourier basis, $A_{\mu\nu}(\bq,\bq')=A_{\mu\nu}(\bq)\delta_{\bq,\bq'}$ with
\begin{equation}
    A_{\mu\nu}(\bm{q}) = \sum_{\bm{R}} A_{\mu\nu}(\bm{0},\bm{R}) e^{-\im\bm{q}\cdot[\bm{R}\,+\,(\bm{c}_\nu-\,\bm{c}_\mu)]}.
    \label{eq:FT-2-index}
\end{equation}
Throughout the main text, we use Miller indices $(hkl)$ to refer to reciprocal space vectors $\bq$, which are given with respect to the conventional cubic cell: $\bq \equiv (hkl) \equiv (2\pi/a_0)[h,k,l].$
We use units where $a_0=1$, so that $q_x = 2\pi h$, $q_y = 2\pi k$, and $q_z = 2\pi l$. For finite $L$, the $L^3$ allowed wavevectors are given by 
\begin{equation}
    \bq = \sum_{m=1}^3 n_m \bm{b}_m/L\,,
    \label{eq:SM-allowed-q}
\end{equation} 
where $n_m\in\{0,\ldots,L-1\}$ and $\bm{b}_m$ are the reciprocal lattice basis vectors, defined by  $\bm{a}_n\cdot\bm{b}_m = 2\pi \delta_{nm}$, 
\begin{equation*}
    \bm{b}_1 = \frac{2\pi}{a_0}[1,1,-1], \quad
    \bm{b}_2 = \frac{2\pi}{a_0}[1,-1,1], \quad
    \bm{b}_3 = \frac{2\pi}{a_0}[-1,1,1]. 
\end{equation*}

\newpage
\section{Analytical Details}
\label{SM-Analytical-Details}

In this section we provide further mathematical details and derivations for the methods and results found in the main text.

\subsection{Polarized Neutron Scattering}
\label{SM-PNS}

\begin{figure}[tb]
	\centering
	\includegraphics[width=0.7\columnwidth]{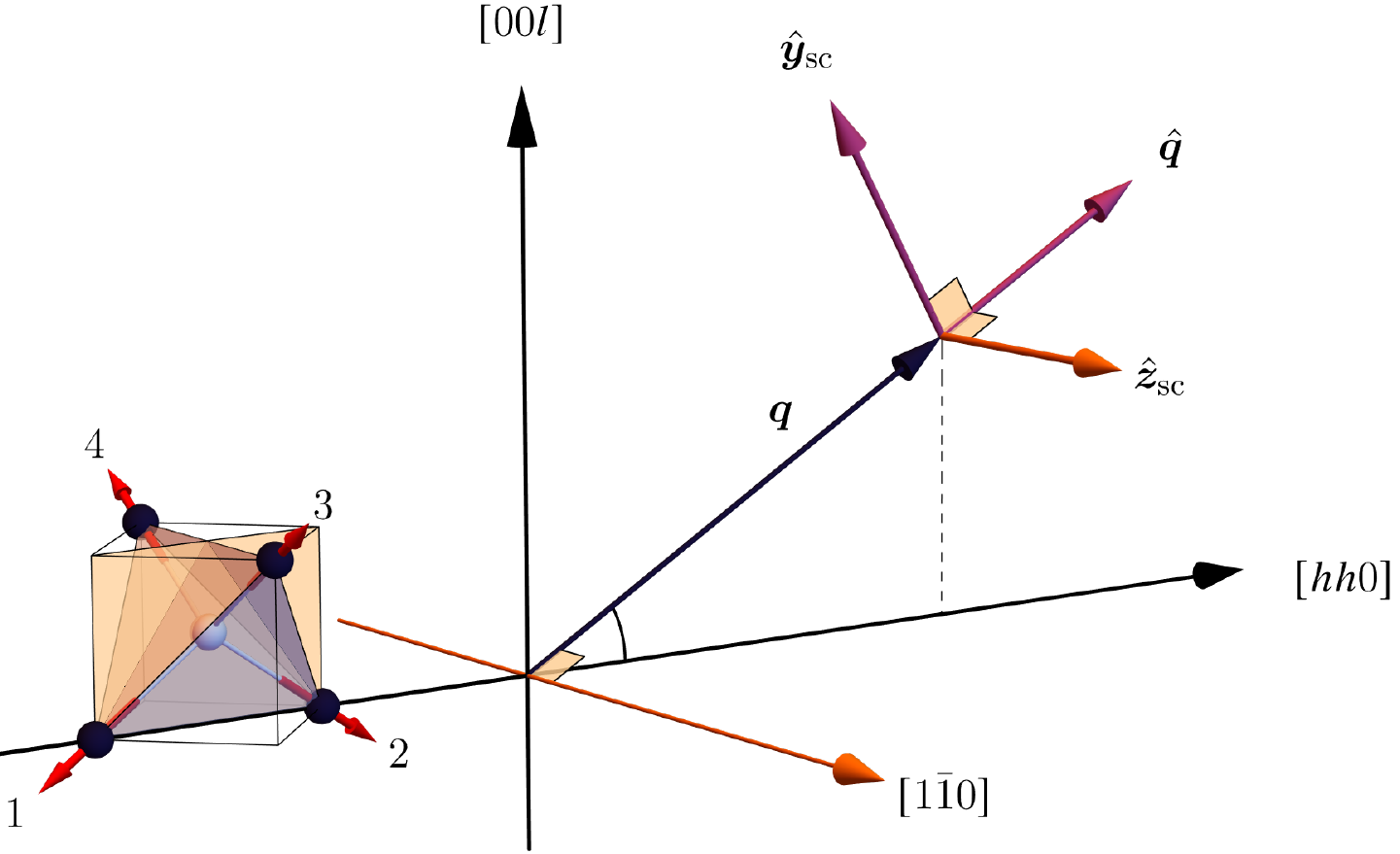}
	\caption{Illustration of the relationship of the polarized neutron scattering basis vectors in the $(hhl)$ scattering plane, along with a tetrahedron of the pyrochlore lattice provided for orientational reference.}
	\label{fig:PNS_diagram}
\end{figure}

The static neutron scattering differential cross section is related to the spin-spin correlation functions through
\begin{equation}
    \left(\frac{\mathrm{d}\sigma}{\mathrm{d}\Omega}\right) \propto \abs{f(\bq)}^2 \sum_{\alpha\beta}\left(\delta^{\alpha\beta}-\hat{q}^\alpha \hat{q}^\beta\right)\sum_{\mu\nu}\langle S_\mu^\alpha(-\bq)S_\nu^\beta(\bq)\rangle\,,
    \label{eq:neutron-cross section}
\end{equation}
where $\bq$ is the scattering wavevector, $\alpha,\beta=x,y,z$ label spin components, and $f(\bq)$ is the electronic form factor of the magnetic ions~\cite{SM-Lovesey1984}. The proportionality constant is a numerical prefactor which is not important for our discussion~\cite{SM-Lovesey1984}. The longitudinal projection factor $\left(\delta^{\alpha\beta}-\hat{q}^\alpha \hat{q}^\beta\right)$ arises from the fact that the neutrons couple to the physical magnetic field which has zero divergence, and so do not ``see'' correlations of spin components along $\bq$. 

Polarization analysis allows for the separation of the remaining two transverse components. Polarizing the neutron spins along a single direction $\uvec{z}_{\mathrm{sc}}$ allows for the definition of an orthonormal basis for each $\bq$ in the plane orthogonal to $\uvec{z}_{\mathrm{sc}}$, with $\uvec{x}_{\mathrm{sc}} \equiv \hat{\bq}$ and $\uvec{y}_{\mathrm{sc}} \equiv \uvec{z}_{\mathrm{sc}}\times\uvec{x}_{\mathrm{sc}}$, as illustrated in \cref{fig:PNS_diagram}. The projection factor in \cref{eq:neutron-cross section} can then be written 
\begin{equation*}
    (\delta^{\alpha\beta}-\hat{q}^\alpha\hat{q}^\beta) = \hat{y}_{\mathrm{sc}}^\alpha\,\hat{y}_{\mathrm{sc}}^\beta + \hat{z}_{\mathrm{sc}}^\alpha\, \hat{z}_{\mathrm{sc}}^\beta\,,
\end{equation*}
so that the cross section \cref{eq:neutron-cross section} splits into a sum of two pieces,
\begin{equation}
    \left(\frac{\mathrm{d}\sigma}{\mathrm{d}\Omega}\right) \propto \vert f(\bq)\vert^2\Big[
    \underbrace{\sum_{\mu\nu}\sum_{\alpha\beta}\hat{y}_{\mathrm{sc}}^\alpha \langle S_\mu^\alpha(-\bq) S_\nu^\beta(\bq)\rangle\hat{y}_{\mathrm{sc}}^\beta}_{\sigma_{\mathrm{SF}}(\bq)}
    +
    \underbrace{\sum_{\mu\nu}\sum_{\alpha\beta}\hat{z}_{\mathrm{sc}}^\alpha \langle S_\mu^\alpha(-\bq) S_\nu^\beta(\bq)\rangle\hat{z}_{\mathrm{sc}}^\beta}_{\sigma_{\mathrm{NSF}}(\bq)}
    \Big]\,.
    \label{eq:SM-SF-NSF-cross-section}
\end{equation}
The first term contains contributions from the spin component orthogonal to the neutron polarization, which induces flipping of the neutron moment and is thus termed the spin-flip (SF) channel, while the second contains spin components parallel to the neutron polarization and is termed the non-spin-flip (NSF) channel. 
We refer to $\SF$ and $\NSF$ as the SF and NSF structure factors, respectively. 
For Ising spins, we replace $S_\mu^\alpha(\bq) = s_\mu(\bq)\hat{z}_\mu^\alpha$ and one obtains the expressions in Eqs.~(2) and (3) in the main text.

\subsection{Large-\textit{N} Correlations}
\label{SM-Large-N-Correlations}

Here we provide a consolidation of the important details of the large-$N$ approximation used to compute the correlation functions in the main text. For further information, see Refs.~\cite{SM-Stanley1968,Garanin1996,Conlon2010,isakov2004}. The Hamiltonian for an isotropic spin model with $N$-component spins $\bm{S}_i$ is given by, 
\begin{equation}
    H = \frac{1}{2} \sum_{ij} V_{ij} \bm{S}_i \cdot \bm{S}_j\,.
    \label{eq:SM-Hamiltonian}
\end{equation}
The correlation functions are computed from
\begin{equation*}
    \mathcal{G}_{ij}^{\alpha\beta} \equiv \langle S_i^\alpha S_j^\beta \rangle = \frac{1}{Z} \int \mathcal{D}\bm{S} \,S_i^\alpha S_j^\beta \,\prod_k\delta(\abs{\bm{S}_k}^2 - N)\,  e^{-\beta H[\bm{S}]}\,,
\end{equation*}
where $\alpha,\beta$ are spin component indices, $Z$ is the partition function, and the integration measure is $\mathcal{D}\bm{S} \equiv \prod_{j}\mathrm{d}^N\!\bm{S}_j$ with the integration domain $S_j^\alpha\in(-\infty,\infty)$. Expanding each delta function as $\delta(\abs{\bm{S}_j}^2-N)\propto\int_{-\infty}^{\infty} \mathrm{d}\phi_j e^{-\im\phi_j(\abs{\bm{S}_j}^2-N)/2}$, the partition function takes the form (up to an overall constant)
\begin{equation*}
    Z \propto \int \mathcal{D}\bm{S} \mathcal{D}\phi \exp(-\frac{1}{2}\sum_{ij}\sum_{\alpha\beta} S_i^{\alpha}\left[ (\im\phi_j \delta_{ij} + \beta V_{ij})\delta_{\alpha\beta}\right]S_j^{\beta}+N\frac{1}{2}\sum_j \im\phi_j)\,.
\end{equation*}
From here, the spins can be integrated out to obtain a theory of only the auxiliary field $\phi$; defining a matrix whose components are functions of the auxiliary field, $[\bar{\mathcal{G}}(\im\phi)^{-1}]_{ij} \equiv (\im\phi_j\delta_{ij} + \beta V_{ij})$, we obtain
\begin{equation*}
    Z \propto \int \mathcal{D}\phi \exp(-N\frac{1}{2}\left[\Tr\log[\bar{\mathcal{G}}(\im\phi)] - \sum_j \im \phi_j\right])\,.
\end{equation*}
This integral can be performed in the limit $N\to \infty$ via saddle point by expanding the exponent about its minimum. The saddle point condition is $[\bar{\mathcal{G}}(\im\phi)]_{jj} = 1$ for each $j$, the solution of which is translationally invariant: $\im\phi_j = \lambda$ for all $j$, where $\lambda$ is a real number. The translation invariance allows the condition to be written in a more convenient form $\Tr[\bar{\mathcal{G}}(\lambda)] = 4L^3$. 

The saddle point approximation of the spin-spin correlation function is then given by $\mathcal{G}_{ij}^{\alpha\beta} = [\bar{\mathcal{G}}(\lambda)]_{ij}\delta_{\alpha\beta}$, i.e.
\begin{equation}
    \langle S_i^\alpha S_j^\beta \rangle_{N\to\infty} = \left[\lambda \openone + \beta V\right]^{-1}_{ij} \delta_{\alpha\beta}\,.
    \label{eq:large-N-corr}
\end{equation}
This expression is equivalent to the correlations of the spherical model, in which the spin length constraints $\abs{\bm{S}_i}^2 = 1$ for each $i$ are replaced by the single constraint $\sum_i \abs{\bm{S}_i}^2 = 4L^3$~\cite{SM-Berlin1952,SM-Stanley1968}. It is also exactly what one would obtain by naively replacing each delta function with a Gaussian, $\delta(\abs{\bm{S}_i}^2-1)\to \exp(-\lambda \abs{\bm{S}_i}^2/2)$, referred to as the self-consistent Gaussian approximation (SCGA) \cite{SM-Conlon2010}. The SCGA partition function is given by
\begin{equation*}
    Z_{\mathrm{SCGA}} = \int \mathcal{D}\bm{S} \exp(-\frac{1}{2}\sum_{ij} (\lambda \delta_{ij} + \beta V_{ij}) \bm{S}_i\cdot\bm{S}_j),
\end{equation*}
yielding an \emph{effective} Hamiltonian
\begin{equation}
    H_{\mathrm{eff}} = \frac{1}{2} \sum_{ij} (\lambda T \delta_{ij} + V_{ij}) \bm{S}_i \cdot \bm{S}_j\,,
    \label{eq:Heff-SCGA}
\end{equation}
where $T$ is temperature (in units with $k_{\mathrm{B}}=1$). 
In the SCGA, the spin length is free to fluctuate, with $(\lambda T + V_{ii})$ controlling the magnitude of the  fluctuations. The saddle point condition determines the value of $\lambda$ such that the variance $\langle \abs{\bm{S}_i}^2\rangle$ is a constant at all temperatures. In this paper we only consider the $N=1$ (Ising) case.

\subsection{Interpretation of \texorpdfstring{$\bm{\lambda}$}{Lambda}}
\label{SM-Interpretation-of-lambda}

Note that in the spin Hamiltonian \cref{eq:SM-Hamiltonian} there is a freedom to include in $V$ diagonal matrix elements $V_{ii}\equiv \varepsilon$ because this simply shifts the energy by a constant since $\abs{\bm{S}_i}^2=\mathrm{const}.$ Doing so changes the value of $\lambda$ to preserve the saddle point condition $\Tr[\lambda\openone + \beta V]^{-1} = 4L^3$; in particular the substitution $V \to V + \varepsilon \openone$ must be compensated by $\lambda \to \lambda - \beta \varepsilon$ in order to maintain the equality. 
From here on and throughout the main text, we adopt the convention of $\varepsilon$ chosen such as to set the minimum eigenvalue of $V$ to zero, making $V$ positive semi-definite.  As we now show, this allows us to interpret $\lambda^{-1}$ as the average thermal occupation value (TOV) of the lowest energy normal modes of the system. 

For conciseness, we use the term ``mode'' to describe any normalized linear combination of the spin variables, $\sum_i \hat{\Psi}_i s_i$, where $\sum_i \vert\hat{\Psi}_i\vert^2 = 1$. Note that the $s_\mu(\bq)$ defined by \cref{eq:SM-s-FT} satisfy this definition. The \emph{normal} modes of the system are the modes defined by the normalized eigenvectors of the interaction matrix $V$, $V \hat{\bm{\eigv}}_n(\bq) = \epsilon_n(\bq) \hat{\bm{\eigv}}_n(\bq)$ where $n\in\{1,2,3,4\}$ (the eigenvectors are indexed by $\bq$ due to translation symmetry). The normal modes are the linear combinations of spin variables $\tilde{s}_n(\bq) \equiv \sum_i [\hat{\eigv}_n(\bq)]_i \,s_i$, in terms of which the effective Hamiltonian \eqref{eq:Heff-SCGA} is
\begin{equation*}
    \beta H_{\mathrm{eff}} = \frac{1}{2}\sum_{\bq,n}(\lambda + \beta \epsilon_n(\bq)) \,\,\abs{\tilde{s}_n(\bq)}^2\,.
\end{equation*}
The large-$N$ normal mode TOV can be read off, 
\begin{equation}
    \left\langle \abs{\tilde{s}_n(\bq)}^2\right\rangle = \frac{1}{\lambda + \beta\epsilon_n(\bq)}\,.
    \label{eq:normal-mode-susceptibility}
\end{equation}
With our convention that $\varepsilon$ is chosen so that  $\min_{\bq,n}[\epsilon_n(\bq)] = 0$, $\lambda$ is always positive and the TOV of the lowest energy normal modes (i.e. those with $\epsilon_n(\bq)=0$) is precisely $\lambda^{-1}$. The saddle point condition can be written $\Tr\mathcal{G}\equiv \sum_{\bq,n} \langle \abs{\tilde{s}_n(\bq)}^2\rangle\equiv \sum_i\langle s_i^2\rangle = 4L^3$, which states that the total TOV of all modes is conserved.

Starting from infinite temperature $(\beta = 0)$, $\mathcal{G}_{ij} = \lim_{\beta \to 0}[\lambda \openone + \beta V]^{-1}_{ij} = \lambda^{-1} \delta_{ij}$, meaning that the spins are fully uncorrelated. The saddle point condition is $4L^3=\lim_{\beta\to 0}\Tr\mathcal{G} = \lambda^{-1}\sum_i \delta_{ii}$, whose solution is $\lambda = 1$. Note from \cref{eq:normal-mode-susceptibility} with $\beta=0$ and $\lambda=1$, every normal mode is singly occupied. As temperature is lowered ($\beta > 0$), normal modes with higher energy depopulate. Conservation of total TOV implies that the lower energy normal mode TOV's must then increase. In particular, the TOV of the lowest energy $\epsilon_n(\bq)=0$ normal modes, $\lambda^{-1}$, must \emph{monotonically} increase (because these modes can never depopulate as temperature is lowered), and so $\lambda$ monotonically decreases. It may happen that for some critical temperature $T_c$, $\lambda\to 0$ as $T\to T_c^+$ (note that $\lambda$ can only reach zero at a non-zero temperature in the thermodynamic limit $L\to\infty$). From \cref{eq:normal-mode-susceptibility} any normal mode with $\epsilon_n(\bq)=0$ becomes \emph{critical} as $\lambda^{-1}\to\infty$, i.e. its susceptibility (proportional to its TOV) diverges, indicating a phase transition to a long-range ordered phase. The specific nature of the symmetry broken ground state spin configuration is dependent on the exact spin length constraint, as it may not be possible to construct a configuration with fixed spin length from only the critical mode eigenvectors of $V$. 

In the case of nearest-neighbor spin ice (NNSI) and extended spin ice (ESI) discussed in the main text, $\lambda$ saturates to $1/2$ at low temperature due to the fact that half of the normal modes (corresponding to the flat band eigenvectors of the respective interaction matrices $V_{\mathrm{NNSI}}$ and $V_{\mathrm{ESI}}$) have $\epsilon_n(\bq) = 0$. All positive energy normal modes (corresponding to dispersive band eigenvectors) are depopulated (TOV$=0$), but since the total TOV of all normal modes is conserved, the flat band normal mode TOVs double, $\langle \vert \tilde{s}_n(\bq) \vert^2 \rangle = 2$.

\subsection{The Interaction Matrix \textit{V}}
\label{SM-The-Interaction-Matrix-V}

The interaction matrix $V$ is defined in terms of the $n$-th neighbor adjacency matrices $A^{(n)}$, 
\begin{equation}
    V_{ij} = \varepsilon\, \delta_{ij} + \sum_n J_n A^{(n)}_{ij}\,,
    \label{eq:SM-V}
\end{equation}
where, again, $\varepsilon$ is chosen to set the lowest eigenvalue of $V$ to zero, $J_n$ are the energies, and the adjacency matrix components are 
\begin{equation*}
    A_{ij}^{(n)} = 
    \begin{cases}
    1           &\quad i,j\text{ are $n$'th nearest-neighbors},\\
    0           &\quad \text{otherwise}.
    \end{cases}
\end{equation*}
For convenience, we give here the Fourier-transformed adjacency matrices $\cramped{A^{(n)}_{\mu\nu}(\bq)}$ (see conventions in \cref{SM-Fourier-Transforms}) for first, second, and third (type $a$) neighbors [see Fig.~1(a) in the main text]. For first and second neighbors, they are 
\begin{equation}
A_{\mu\nu}^{(1)} (\bq) 
= 2 \,
\begin{pmatrix}
0 & c_{xy} & c_{xz} & c_{yz} \\
c_{xy} & 0 & c_{y\overline{z}} & c_{x\overline{z}} \\
c_{xz} & c_{y\overline{z}} & 0 & c_{x\overline{y}}\\
c_{yz} & c_{x\overline{z}} & c_{x\overline{y}} & 0
\end{pmatrix} 
\,,
\qquad 
A_{\mu\nu}^{(2)} (\bq) 
= 4 \,
\begin{pmatrix}
0 & c_{zz} \, c_{x\overline{y}} & c_{yy} \, c_{x\overline{z}} & c_{xx} \, c_{y\overline{z}} \\
c_{zz} \, c_{x\overline{y}} & 0 & c_{xx} \, c_{yz} & c_{yy} \, c_{xz} \\
c_{yy} \, c_{x\overline{z}} & c_{xx} \, c_{yz} & 0 & c_{zz} \, c_{xy} \\
c_{xx} \, c_{y\overline{z}} & c_{yy} \, c_{xz} & c_{zz} \, c_{xy} & 0
\end{pmatrix} \,,\\
\label{eq:interaction_matrix_1_2}
\end{equation}
with $\cramped{c_{ab} \equiv \cos(\frac{q_a + q_b}{4})}$, $\cramped{c_{a\overline{b}} \equiv \cos(\frac{q_a - q_b}{4})}$, where $a, b$ represent the Cartesian coordinates $x,y,z$, as appropriate. For type-$a$ third neighbors, 
\begin{equation}
\begin{split}
A_{\mu\nu}^{(3a)} (\bq) 
&= 2 \,
\begin{pmatrix}
C_{\scriptscriptstyle +\,+\,+} & 0 & 0 & 0 \\
0 & C_{\scriptscriptstyle +\,-\,-} & 0 & 0 \\
0 & 0 & C_{\scriptscriptstyle -\,-\,+} & 0 \\
0 & 0 & 0 & C_{\scriptscriptstyle -\,+\,-}
\end{pmatrix}\,,
\label{eq:interaction_matrix_3a}
\end{split}
\end{equation}
with $C_{\scriptscriptstyle +\,-\,-} =  \cos(\frac{q_{x}+q_{y}}{2}) + \cos(\frac{q_{y}-q_{z}}{2}) + \cos(\frac{q_{z}-q_{x}}{2})$, 
and so on. 

In the main text, we discuss the nearest-neighbor and extended spin ice models. The interaction matrix of NNSI is given by \cref{eq:SM-V} with only nearest-neighbor $J_1$ interactions, and $\varepsilon = 2J_1$ to shift the minimum eigenvalue to zero,
\begin{equation}
    V_{\mathrm{NNSI}} = 2J_1 \openone + J_1 A^{(1)}\,.
    \label{eq:SM-VNNSI}
\end{equation}
The interaction matrix of ESI includes nearest-neighbor $J_1$, as well as equal $J_2=J_{3a}\equiv J'$ interactions, with $\varepsilon=2(J_1-J')$ to shift the minimum eigenvalue to zero, 
\begin{equation}
    V_{\mathrm{ESI}}  = 2(J_1-J')\openone + J_1 A^{(1)} + J'(A^{(2)} + A^{(3a)})\,.
    \label{eq:SM-VESI}
\end{equation}

\subsection{Flat Band Eigenvector \texorpdfstring{$\hat{\bm{\Omega}}$}{Omega-hat} of the Adjacency Matrix}
\label{SM-Flat-Band-Eigenvector}

The Fourier transformed pyrochlore adjacency matrix, \cref{eq:interaction_matrix_1_2}, for $\bq\in(hhl)$ is 
\begin{equation*}
    A_{\mu\nu}^{(1)} (\bq) 
    = 2 \,
    \begin{pmatrix}
    0 & c_{hh} & c_{hl} & c_{hl} \\
    c_{hh} & 0 & c_{h\overline{l}} & c_{h\overline{l}} \\
    c_{hl} & c_{h\overline{l}} & 0 & 1\\
    c_{hl} & c_{h\overline{l}} & 1 & 0
    \end{pmatrix} 
\end{equation*}
It is easy to verify that $\hat{\bm{\Omega}}$ with components $\hat{\Omega}_{\mu}=(0,0,1,-1)/\sqrt{2}$ is an eigenvector of this matrix with eigenvalue $-2$ independent of $h$ or $l$, which is precisely the eigenvalue of the flat bands of $A^{(1)}$.

\subsection{Flat NSF of NNSI and ESI}
\label{SM-Flat-NSF}

\subsubsection{ESI Interaction Matrix}
\label{SM-ESI-Interaction-Matrix}

The interaction matrix for ESI, \cref{eq:SM-VESI} (which includes NNSI when $J'=0$), simplifies significantly when $\bq\in(hhl)$:
\begin{equation}
[V_{\mathrm{ESI}}]_{\mu\nu}(\bq) = 2(J_1-J')\delta_{\mu\nu} + J_1 A_{\mu\nu}^{(1)}(\bq) + J'\left( A_{\mu\nu}^{(2)}(\bq) + A_{\mu\nu}^{(3a)}(\bq)\right) = 2 
\begin{pmatrix*}[l]
a_+ & b & c_+ & c_+ \\
b & a_- & c_- & c_- \\
c_+ & c_- & d & d \\
c_+ & c_- & d & d
\end{pmatrix*}\,,
\label{eq:V-ESI}
\end{equation}
where 
\begin{align*}
    a_{\pm}            &= J_1 + J'\left[\cos(2\pi h)+2\cos(\pi(h\pm l))-1\right]\,, \nonumber\\
    b_{\hphantom{\pm}} &= J_1\cos(\pi h) + 2J' \cos(\pi l)\,, \nonumber\\
    c_{\pm}            &= J_1\cos(\pi(h\pm l)/2) + 2J'\cos(\pi h)\cos(\pi(h\mp l)/2)\,, \nonumber \\
    d_{\hphantom{\pm}} &= J_1 + 2 J'\cos(\pi h)\cos(\pi l)\,.
\end{align*}
Importantly, note that the third and fourth columns of \cref{eq:V-ESI} are identical for \emph{all} $\bq\in(hhl)$, so that $\hat{\bm{\Omega}}$ is again an eigenvector of this matrix, with eigenvalue zero.

\subsubsection{Flat NSF}
\label{SM-ESI-Flat-NSF}

Using the conventions \eqref{eq:local_z_axes}, the non-spin-flip (NSF) projection factors $(\uvec{z}_{\mathrm{s}}\cdot\uvec{z}_\mu)$ for scattering in the $(hhl)$ plane with $\uvec{z}_{\mathrm{s}} = [-1,1,0]/\sqrt{2}$ are given by
\begin{equation}
\nsfvec_\mu \equiv \uvec{z}_{\mu} \cdot \uvec{z}_{\mathrm{s}} =
\begin{cases}
0 &\quad \mu = 1, \\
0 &\quad \mu = 2, \\
+\sqrt{2/3} &\quad \mu = 3,\\
-\sqrt{2/3} &\quad \mu = 4.\
\end{cases}
\label{eq:nsf_projection_factors-hhl}
\end{equation}
which is $\sqrt{4/3}\,\hat{\nsfvec}$. On the other hand, the spin-flip (SF) projection factors $(\uvec{y}_{\mathrm{s}}\cdot\uvec{z}_{\mu})=(\uvec{z}_{\mathrm{s}}\times \hat{\bq})\cdot\uvec{z}_{\mu}$ are all $\bq$-dependent and non-zero. 

In the large-$N$ approximation, the correlation matrix is given by $\mathcal{G}(\bq) = [\lambda\openone+\beta V_{\mathrm{ESI}}(\bq)]^{-1}$. For $\bq\in(hhl)$, $\hat{\bm{\nsfvec}}$ is then an eigenvector of $\mathcal{G}(\bq)$ with eigenvalue $\lambda^{-1}$. Utilizing the expression $\NSF(\bq) = \vert\bm{\nsfvec}\vert^2 \, \hat{\nsfvec}_\mu \mathcal{G}_{\mu\nu}(\bq) \hat{\nsfvec}_{\nu}$ yields $\NSF(\bq)=\abs{\bm{\nsfvec}}^2/\lambda=4/3\lambda$, as reported in the main text.

\subsection{SF and NSF Correlation Functions in Terms of \textbf{\textit{B}} Field}
\label{SM-SF-NSF-B}

Utilizing the expression for $\NSF$ given in \cref{eq:SM-SF-NSF-cross-section}, we can write
\begin{align*}
    \NSF(\bq) = \sum_{\mu\nu} \langle [S_\mu^\alpha(-\bq)\hat{z}_s^\alpha][S_\nu^\beta(\bq)\hat{z}_s^\beta]\rangle 
    = \left< \left[\sum_\mu\bm{S}_\mu(-\bq)\cdot\uvec{z}_{\mathrm{sc}}\right]
    \left[\sum_\nu\bm{S}_\nu(\bq)\cdot\uvec{z}_{\mathrm{sc}}\right] 
    \right> 
    = \left< \left\vert\sum_\mu\bm{S}_\mu(\bq)\cdot\uvec{z}_{\mathrm{sc}}\right\vert^2 \right> 
    = \frac{4}{3}\left< \left\vert\bm{B}(\bq)\cdot\uvec{z}_{\mathrm{sc}}\right\vert^2 \right>\,, 
\end{align*}
where $\bm{B}(\bq)=\sqrt{3/4}\,\sum_\mu \bm{S}_\mu(\bq)$ and $\bm{S}_\mu(\bq)\equiv s_\mu(\bq) \,\uvec{z}_\mu$. Similarly, we obtain $\SF=\frac{4}{3}\langle \vert \bm{B}(\bq)\cdot\uvec{y}_{\mathrm{sc}}\vert\rangle$.

\subsection{Direct Space (\textit{hhl}) NSF Correlations}
\label{SM-Direct-Space-NSF}

\subsubsection{NSF as Chain-Chain Correlator}
\label{SM-NSF-chain-chain}

In \cref{SM-SF-NSF-B}, we showed that
\begin{align}
    \NSF(\bq) &= \left< \left| \sum_{\mu}\bm{S}_\mu(\bq) \cdot \uvec{z}_{\mathrm{sc}}\right|^2\right> \,.
    \label{eq:SM-NSF-S}
\end{align}
For $\bq\in (hhl)$ and $\uvec{z}_{\mathrm{sc}}\parallel [\bar{1}10]$, $\bm{S}_1(\bq)\cdot\uvec{z}_{\mathrm{sc}} = \bm{S}_2(\bq)\cdot\uvec{z}_{\mathrm{sc}} = 0$. The remaining two sublattices 3 and 4 lie along the one-dimensional $[\bar{1} 1 0]$ $\alpha$-chains [see Fig.1(a) in main text], which are perpendicular to the direct-space $[hhl]$ plane. Expanding out the above expression we then obtain
\begin{align*}
    \sum_{\mu} \bm{S}_\mu(\bq) \cdot \uvec{z}_{\mathrm{sc}} &= \bm{S}_3(\bq)\cdot\uvec{z}_{\mathrm{sc}} + \bm{S}_4(\bq)\cdot\uvec{z}_{\mathrm{sc}}\,,\\
    &= \frac{1}{\sqrt{L^3}}\sum_{\bm{r}}\left[ \bm{S}_3(\bm{r})e^{-i\bq\cdot\bm{c}_3}+ \bm{S}_4(\bm{r})e^{-i\bq\cdot\bm{c}_4}\right]\cdot\uvec{z}_{\mathrm{sc}} \, e^{-i\bq\cdot\bm{r}}\,,\\
    &= \frac{1}{\sqrt{L^3}}\sum_{\bm{r}}\left[ \bm{S}_3(\bm{r})+ \bm{S}_4(\bm{r})e^{-i\bq\cdot(\bm{c}_4-\bm{c}_3)}\right]\cdot\uvec{z}_{\mathrm{sc}}\, e^{-i\bq\cdot\bm{r}}e^{-i\bq\cdot\bm{c}_3}\,.
\intertext{Note that for $\bq\in(hhl)$, $\bq\cdot(\bm{c}_4-\bm{c}_3) = 0$. Since each spin on sublattice $3$ or $4$ belongs to a single $\alpha$-chain, we can rewrite the sum over $L^3$ unit cells with positions $\bm{r}$ as a sum over the $L^2$ $\alpha$-chains and the $L$ unit cells in each chain,}
    &= e^{-i\bq\cdot\bm{c}_3} \frac{1}{\sqrt{L^2}}\sum_{\alpha}\frac{1}{\sqrt{L}}\sum_{\bm{r}\in\alpha}\left[ \bm{S}_3(\bm{r})\cdot\uvec{z}_{\mathrm{sc}}+ \bm{S}_4(\bm{r})\cdot\uvec{z}_{\mathrm{sc}}\right]\, e^{-i\bq\cdot\bm{r}}\,.
\end{align*}
For a given chain $\alpha$, $\bq\cdot\bm{r}$ is equivalent to $\bq\cdot\bm{R}_\alpha$, where $\bm{R}_\alpha$ is the orthogonal separation of chain $\alpha$ from the origin, i.e. the projection of $\bm{r}$ to the direct-space $[hhl]$ plane (since $\bq\in(hhl)$). The overall phase factor is irrelevant since \cref{eq:SM-NSF-S} only utilizes the modulus square of this quantity. Define 
\begin{equation*}
    \mathsf{P}_\alpha \equiv \frac{1}{\sqrt{L}}\sum_{\bm{r}\in\alpha}[\bm{S}_3(\bm{r})\cdot\uvec{z}_{\mathrm{sc}} + \bm{S}_4(\bm{r})\cdot\uvec{z}_{\mathrm{sc}}] \equiv \frac{1}{\sqrt{L}}\sum_{i\in \alpha} \bm{S}_i\cdot\uvec{z}_{\mathrm{sc}}\,,
\end{equation*}
which is the (normalized) total spin of the chain projected along $\uvec{z}_{\mathrm{sc}}$. Then 
\begin{align*}
    \sum_{\mu} \bm{S}_\mu(\bq) \cdot \uvec{z}_{\mathrm{sc}} &= e^{-i\bq\cdot\bm{c}_3} \frac{1}{\sqrt{L^2}}\sum_{\alpha}\mathsf{P}_\alpha\, e^{-i\bq\cdot\bm{R}_\alpha}\equiv e^{-i\bq\cdot\bm{c}_3} \mathsf{P}(\bq)\,,
\end{align*}
and plugging back in to \cref{eq:SM-NSF-S}, we obtain
\begin{equation*}
    \NSF(\bq) = \langle \vert \mathsf{P}(\bq) \vert^2 \rangle = \frac{1}{L^2} \sum_{\alpha,\alpha'} \langle \mathsf{P}_\alpha \mathsf{P}_{\alpha'} \rangle e^{-i\bq\cdot(\bm{R}_{\alpha'} - \bm{R}_\alpha)}\,,
\end{equation*}
thus demonstrating that in this plane $\NSF$ is the Fourier-transformed chain-chain correlation function $\langle \mathsf{P}_\alpha\mathsf{P}_{\alpha'}\rangle$.

\subsubsection{Connection Between Chains and Flat Band Eigenvectors}
\label{SM-chains-and-eigenvector}

It was shown in Ref.~\cite{SM-bergman2008} that the pyrochlore adjacency matrix $A^{(1)}$ has flat band eigenvectors whose direct-space components are zero everywhere except along a single chain, along which it alternates in sign. In particular, for each $\alpha$-chain, we define a normalized vector $\hat{\bm{\Psi}}_\alpha$ $(\sum_i [\hat{\Psi}_\alpha]_i ^2 = 1)$ with direct-space components
\begin{equation*}
    [\hat{\Psi}_\alpha]_i = \begin{cases}
        +1/\sqrt{2L} &\quad \text{site $i$ is on chain $\alpha$ and sublattice 3},\\
        -1/\sqrt{2L} &\quad \text{site $i$ is on chain $\alpha$ and sublattice 4},\\
        0 &\quad \text{otherwise}.
    \end{cases}
\end{equation*}
This is a flat band eigenvector of $A^{(1)}$ with eigenvalue $-2$, which we write in compact form as
\begin{equation}
    [\hat{\Psi}_\alpha]_i = \frac{1}{\sqrt{L}} \hat{\Omega}_{\mu(i)} \delta_{i\in \alpha}\,,
    \label{eq:SM-Psi-alpha}
\end{equation}
where $\mu(i)$ is the sublattice of site $i$, $\delta_{i\in\alpha}$ is a Kronecker delta which is 1 if site $i$ is in chain $\alpha$ and zero otherwise, and $\hat{\Omega}_\mu=(0,0,1,-1)/\sqrt{2}$. Next, for $\bq\in(hhl)$, we define
\begin{equation}
    \hat{\bm{\Psi}}(\bq) \equiv \frac{1}{\sqrt{L^2}}\sum_\alpha e^{-i\bq\cdot\bm{R}_\alpha} \hat{\bm{\Psi}}_\alpha\,.
    \label{eq:SM-Psi-q}
\end{equation}
This is a (normalized) linear combination of the $\hat{\bm{\Psi}}_\alpha$, and is therefore again a flat band eigenvector of $A^{(1)}$. This vector represents a plane wave, for which the $\alpha$-chains lie in the planes of constant phase. Finally, the mode probed by the $(hhl)$ NSF is
\begin{equation*}
    \hat{\Omega}_\mu s_\mu(\bq) = \frac{1}{\sqrt{L^3}} \sum_i \hat{\Omega}_{\mu(i)} e^{-i\bq\cdot\bm{r}_i} s_i = \frac{1}{\sqrt{L^2}} \sum_\alpha e^{-i\bq\cdot\bm{R}_\alpha} \left(\sum_{i\in\alpha} \frac{1}{\sqrt{L}} \hat{\Omega}_{\mu(i)} s_i \right) = \frac{1}{\sqrt{L^2}} \sum_\alpha e^{-\bq\cdot\bm{R}_\alpha} \sum_i [\hat{\Psi}_\alpha]_i s_i = \sum_i [\hat{\Psi}(\bq)]_i s_i.
\end{equation*}
In the first equality, we simply expanded the Fourier transform in terms of the direct-space $s_i$. In the second equality we rearranged the sum over all sites $i$ into a sum over $\alpha$-chains and a sum over each spin in a chain, utilizing the fact that $\hat{\Omega}_1=\hat{\Omega}_2=0$, and $\bq\cdot\bm{r}_i = \bq\cdot\bm{R}_\alpha$ for $\bq\in(hhl)$ and $\mu(i)=3,4$. In the third equality we used \cref{eq:SM-Psi-alpha}, and in the final equality we used \cref{eq:SM-Psi-q}.

\subsection{Perturbative Expansion}
\label{SM-Perturbative-Expansion}

We split the interaction matrix into two pieces,
\begin{equation*}
    V = \Big[\varepsilon_0 \openone + J_1 A^{(1)} + J'\Big(A^{(2)} + A^{(3a)}\Big)\Big] + \Big[ \varepsilon_p \openone + \sum_n J_n' A^{(n)}\Big] \equiv V_0 + V_p\,,
\end{equation*}
where $\max(J_n')\equiv J_p \ll J_1$. $\varepsilon_0$ is chosen so that $V_0$ has minimum eigenvalue zero, while $\varepsilon_p$ is chosen such that the entire interaction matrix $V$ has minimum eigenvalue zero. Taking \cref{eq:large-N-corr} (with $N=1$) and splitting $\lambda = \lambda_0 + \lambda_p$, we rewrite the equation as
\begin{equation*}
    \mathcal{G}^{-1} = (\lambda_0 \openone + \beta V_0) + (\lambda_p \openone + \beta V_p) \equiv \mathcal{G}_0^{-1} + \Sigma\,,
\end{equation*}
where $\mathcal{G}_0=[\lambda_0\openone + \beta V_0]^{-1}$ is the unperturbed correlation matrix, with $\lambda_0$ determined by the unperturbed saddle point condition $\Tr[\mathcal{G}_0]=4L^3$. The perturbed correlation matrix is then given by
\begin{equation*}
    \mathcal{G} = [\mathcal{G}_0^{-1} + \Sigma]^{-1} =  [\openone + \mathcal{G}_0\Sigma ]^{-1}\mathcal{G}_0\,,
\end{equation*}
which may be more familiar to the reader in the form of the Dyson equation, $\mathcal{G} = \mathcal{G}_0 - \mathcal{G}_0 \Sigma \mathcal{G}$. The matrix inverse may be expanded perturbatively as a geometric series so long as the eigenvalues of $\mathcal{G}_0\Sigma$ (proportional to $J_p$) are bounded between -1 and 1, to obtain 
\begin{equation}
    \mathcal{G} = \mathcal{G}_0 - \mathcal{G}_0 \Sigma \mathcal{G}_0 + \mathcal{G}_0 \Sigma \mathcal{G}_0 \Sigma \mathcal{G}_0  + \cdots\,.
    \label{eq:geometric-series}
\end{equation}
$\lambda_p$ can be computed order by order from the condition that $\Tr[\mathcal{G}] = 4L^3$ which, at first order, reads
\begin{equation*}
    \Tr[\mathcal{G}_0] - \Tr[\mathcal{G}_0\Sigma\mathcal{G}_0] = 4L^3\,.
\end{equation*}
However, $\Tr[\mathcal{G}_0] = 4L^3$ by definition of $\lambda_0$, so the condition is
\begin{equation*}
    0= \Tr[\mathcal{G}_0\Sigma\mathcal{G}_0] = \Tr[\mathcal{G}_0 (\lambda_p \openone + \beta V_p) \mathcal{G}_0] =
    \lambda_p\Tr[\mathcal{G}_0 \mathcal{G}_0] + \Tr[\mathcal{G}_0(\beta V_p) \mathcal{G}_0]\,,
\end{equation*}
the solution of which is
\begin{equation}
    \lambda_p = - \frac{\Tr[\mathcal{G}_0(\beta V_p)\mathcal{G}_0]}{\Tr[\mathcal{G}_0^2]}\,.
    \label{eq:SM-lambda-1}
\end{equation}
To second order, $\lambda_p$ is determined by the condition $\Tr[-\mathcal{G}_0 \Sigma \mathcal{G}_0 + \mathcal{G}_0 \Sigma \mathcal{G}_0 \Sigma \mathcal{G}_0] = 0$, and so on.

\subsection{Dipolar Interactions}
\label{SM-Dipolar-Interaction}

The dipolar interaction Hamiltonian can be written (assuming Ising moments) as
\begin{equation*}
    H_{\mathrm{dip}} = \sum_{i<j} s_i \left( \frac{\mu_0 m^2}{4\pi}\frac{\uvec{z}_i \cdot \uvec{z}_j - 3(\uvec{z}_i\cdot\uvec{r}_{ij})(\uvec{z}_j\cdot\uvec{r}_{ij})}{\abs{\bm{r}_{ij}}^3}\right) s_j \equiv \frac{1}{2}\sum_{ij} s_i (J_{\mathrm{dip}} \mathcal{D}_{ij}) s_j
\end{equation*}
where $m$ is the magnetic moment of a single ion, and $J_{\mathrm{dip}}\mathcal{D}_{ij}$ is given by the quantity in parentheses, with $\mathcal{D}_{ii} = 0$ and $\mathcal{D}_{ij}=1$ for nearest-neighbors. This corresponds to $J_{\mathrm{dip}}=(5/3)\mu_0 m^2/4\pi a_{\mathrm{nn}}^3$, where $a_{\mathrm{nn}}=a_0/\sqrt{8}$ is the nearest-neighbor distance on the pyrochlore lattice~\cite{SM-Hertog2000}.


\newpage
\section{Numerical Details}
\label{SM-Numerical-Details}

In this section, we provide details on numerical calculations used to obtain results reported in the main text and in this
Supplemental Material.

\subsection{Monte Carlo}
\label{SM-Monte-Carlo}

Our Monte Carlo simulations were performed using a cubic unit cell with 16 pyrochlore sites per cell.
The system was simulated with $6^3$ cubic unit cells,  containing $N_s= 16\times 6^3 = 3456$ Ising spins, subject to periodic boundary conditions. We employed a standard single spin-flip Metropolis algorithm, supplemented with non-local loop updates at low temperatures to prevent spin freezing~\cite{SM-melko2004}. At each temperature, $4\times 10^5$ iterations were used for thermalization. $5\times 10^4$ sample configurations were used for calculation of thermal averages, with between 60-160 Monte Carlo iterations between each sample, with more at lower temperatures. One iteration attempts a total of $N_s$ single spin-flips and loop updates, with more loop updates at lower temperatures. Loop updates were only used for $T/J_1 \leq 1$.

\begin{figure}[t]
	\centering
	\includegraphics[width=0.9\columnwidth]{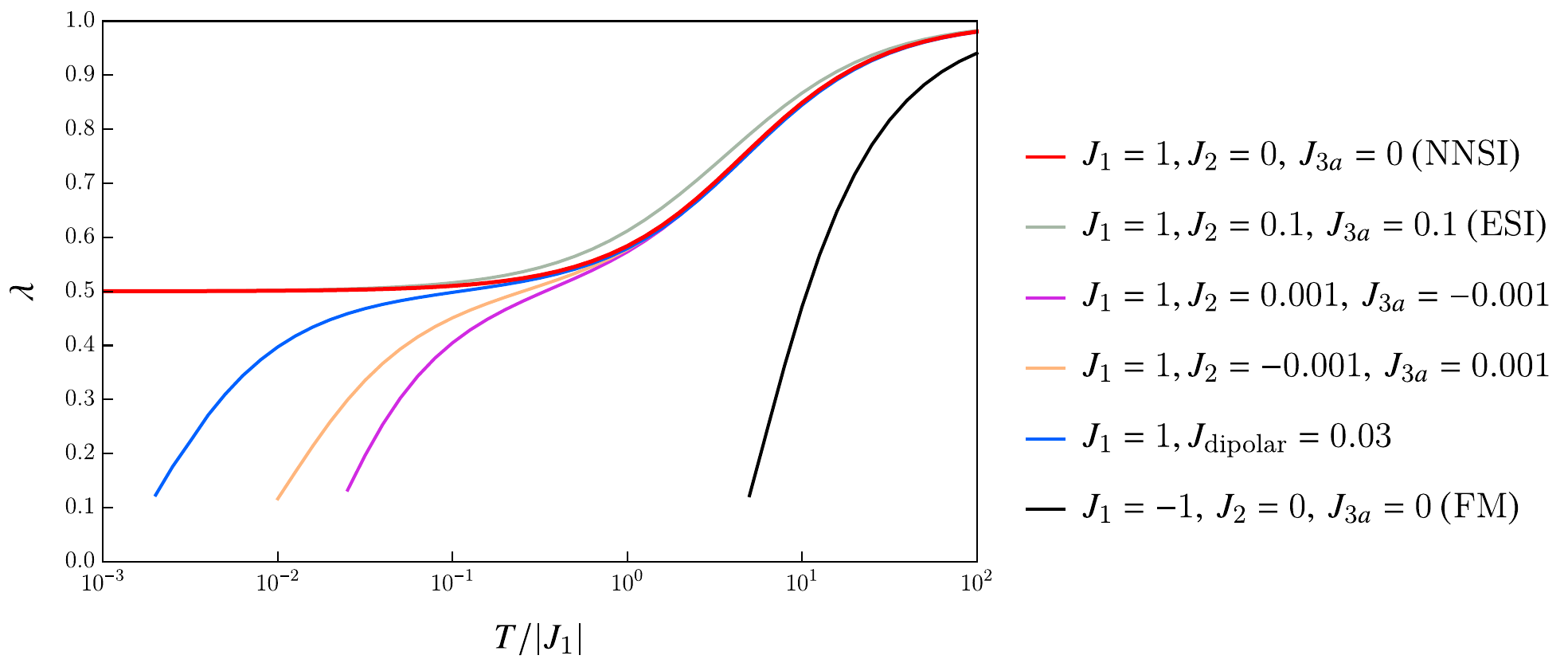}
	\caption{The value of $\lambda$ as a function of $T/\abs{J_1}$ calculated from the saddle point condition $\Tr[\mathcal{G}]=4L^3$ with $L=12$. In all cases, $\lambda = 1$ at high temperature, indicative of the uncorrelated paramagnetic phase [\cref{SM-Interpretation-of-lambda}]. For both NNSI ($J_2 = J_{3a}=0$) and ESI (with $J_2=J_{3a}=0.1$), $\lambda$ approaches $0.5$ at low temperature, indicative of the extensive ground state degeneracy and corresponding Coulomb phase [\cref{SM-Interpretation-of-lambda}]. Off of the ESI line, the ground state degeneracy is lifted, and $\lambda$ tends towards zero at sufficiently low temperature, also true for the ferromagnetic Ising model [\cref{SM-NSF-FM}] and the dipolar case [\cref{SM-Dipolar-Interaction} and \cref{SM-Dipolar-Matrix}]. $\lambda = 0$ at a non-zero temperature indicates a phase transition to a long-range-ordered phase [\cref{SM-Interpretation-of-lambda}]. In the dipolar case in particular, the lifting of the degeneracy is weak, so the phase transition occurs at a low temperature compared to the interaction energy scale $J_{\mathrm{dip}}$~\cite{SM-Melko2001}.}
	\label{fig:lambda-temp-dependence}
\end{figure}

\subsection{Large-\textit{N}}
\label{SM-Large-N-Numerics}

In order to calculate the large-$N$ correlation matrix $\mathcal{G}_{ij}$ at a given temperature, we need to determine $\lambda$ from the saddle point condition $\Tr\,[\lambda \openone + \beta V]^{-1} = 4L^3$. The trace is straightforward to perform as a sum over wave vectors and sublattices, 
\begin{equation*}
    \Tr\,[\lambda \openone + \beta V]^{-1} = \sum_{\bq}\sum_{\mu}[\lambda \openone_{4\times 4} + \beta V(\bq)]^{-1}_{\mu\mu}\,,
\end{equation*} 
where $V_{\mu\nu}(\bq) = \varepsilon \delta_{\mu\nu} + \sum_n J_n A^{(n)}_{\mu\nu}(\bq)$ (see \cref{SM-ESI-Interaction-Matrix} for the definition of the Fourier transformed adjacency matrices). In this form, the saddle point equation can be quickly solved numerically using a root finding algorithm (e.g. Newton descent), in the form
\begin{equation*}
    \left[\frac{1}{4L^3}\sum_{\bq}\sum_{\mu}[\lambda \openone + \beta V(\bq)]^{-1}_{\mu\mu}\right] - 1 = 0\,,
\end{equation*}
for $L\sim 10$, with the $L^3$ allowed $\bq$ vectors given by \cref{eq:SM-allowed-q}. The temperature evolution of $\lambda$ for the various cases considered in the main text and SM are shown in \cref{fig:lambda-temp-dependence}.

\subsection{Experimental Data Color Map Digitization}
\label{SM-Color-Map}

In Fig. 3 of the main text we plotted data from Ref.~\cite{SM-fennell2009}, which was extracted from their Fig.~2B via the following method. First the figure was converted to an array of color values encoded in the CIE 1976 $L^*a^*b^*$ color space~\cite{SM-Luo2015}. Each pixel in the array is then converted to an intensity value by matching it to the corresponding color in the color bar. The matching is performed by minimizing the Euclidean distance between their ($L^*,a^*,b^*$) values.

\subsection{Dipolar Interaction Matrix}
\label{SM-Dipolar-Matrix}

\begin{figure}[t]
    \centering
    \includegraphics[width=0.65\columnwidth]{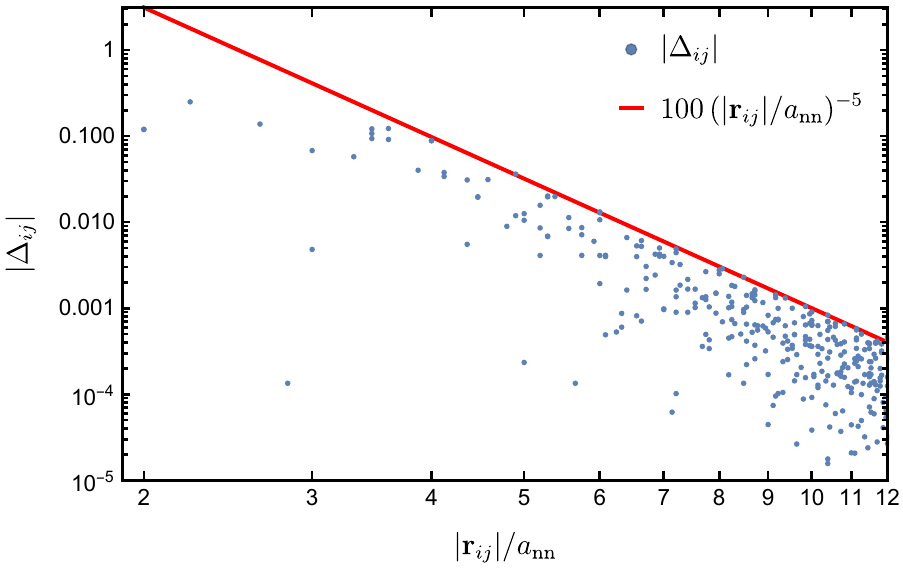}
    \caption{The decay of the correction term magnitude $\abs{\Delta}_{ij}$, computed using \cref{eq:SM-Delta-ij}, as a function of the distance between sites $i$ and $j$, $\abs{\bm{r}_{ij}}$, in units of the nearest-neighbor separation on the pyrochlore lattice $a_{\mathrm{nn}}$. The red line is a guide to the eye, demonstrating the corrections decay as $\abs{\bm{r}_{ij}}^{-5}$, indicating that the long-range $\abs{\bm{r}_{ij}}^{-3}$ portion of the dipolar interaction is fully captured by $c\mathcal{P}_{ij}$.}
    \label{fig:SM-Delta-decay}
\end{figure}

Using the method of \cite{SM-isakov2005}, the dipolar interaction matrix [see \cref{SM-Dipolar-Interaction}] can be written as $\mathcal{D}_{ij} = c\mathcal{P}_{ij} + \Delta_{ij}$, where $\mathcal{P}$ is the projection matrix to the \emph{dispersive} band eigenspace of the nearest-neighbor adjacency matrix $A^{(1)}$. Our definition of $\mathcal{D}$ given in \cref{SM-Dipolar-Interaction} is $(3/5)$ times theirs, and so in our case $c=8\pi/5$. The projection matrix $\mathcal{P}$ can be obtained using the method of \cite{SM-Henley2005}, defining
\begin{equation*}
    \uvec{u}_{1} = \frac{1}{4}[ 1, 1, 1]\,, \quad
    \uvec{u}_{2} = \frac{1}{4}[-1,-1, 1]\,, \quad
    \uvec{u}_{3} = \frac{1}{4}[-1, 1,-1]\,, \quad
    \uvec{u}_{4} = \frac{1}{4}[ 1,-1,-1]\,,
\end{equation*}
and
\begin{equation}
    E(\bq) = \begin{pmatrix}
    e^{\im\bq\cdot\uvec{u}_1/2} & e^{-\im\bq\cdot\uvec{u}_1/2} \\
    e^{\im\bq\cdot\uvec{u}_2/2} & e^{-\im\bq\cdot\uvec{u}_2/2} \\
    e^{\im\bq\cdot\uvec{u}_3/2} & e^{-\im\bq\cdot\uvec{u}_3/2} \\
    e^{\im\bq\cdot\uvec{u}_4/2} & e^{-\im\bq\cdot\uvec{u}_4/2}
    \end{pmatrix}\,,
\end{equation}
the projector is given in reciprocal space by~\cite{SM-Henley2005}
\begin{equation*}
    \mathcal{P}(\bq) = E(\bq) [E(\bq)^\dagger E(\bq)]^{-1} E(\bq)^\dagger\,.
\end{equation*}
For convenience, we provide the exact expression obtained from this equation: defining
\begin{align*}
F(\bq) &\equiv -3 + \cos(\frac{q_x}{2})\cos(\frac{q_y}{2}) + \cos(\frac{q_y}{2})\cos(\frac{q_z}{2}) + \cos(\frac{q_z}{2})\cos(\frac{q_x}{2}) \,, \\
f_{abc}^{\pm}(\bq) &\equiv \cos(\frac{q_a}{4})\cos(\frac{q_b}{4})\left[\cos(\frac{q_c}{2})-1\right] 
\pm 
\sin(\frac{q_a}{4})\sin(\frac{q_b}{4})\left[\cos(\frac{q_c}{2})+1\right]\,, \\
g_{\pm\pm\pm}(\bq) &= \frac{1}{2}\left[F(\bq) \pm \sin(\frac{q_x}{2})\sin(\frac{q_y}{2}) \pm \sin(\frac{q_y}{2})\sin(\frac{q_z}{2}) \pm \sin(\frac{q_z}{2})\sin(\frac{q_x}{2})\right] \, ,
\end{align*}
the projector is then given by
\begin{equation*}
    \mathcal{P}_{\mu\nu}(\bq) = \frac{1}{F(\bq)}
    \begin{pmatrix}
    g_{---} & f_{xyz}^+ & f_{zxy}^+ & f_{yzx}^+  \\
    f_{xyz}^+ & g_{-++} & f_{yzx}^- & f_{zxy}^-  \\
    f_{zxy}^+ & f_{yzx}^- & g_{++-} & f_{xyz}^-  \\
    f_{yzx}^+ & f_{zxy}^- & f_{xyz}^- & g_{+-+} 
    \end{pmatrix}  \, .
\end{equation*}
Note that this is singular at $\bq=\bm{0}$, where one dispersive band touches the flat bands. 

To obtain the correction terms $\Delta_{ij}$, we compute
\begin{equation}
    \Delta_{ij} = \mathcal{D}_{ij} - c\mathcal{P}_{ij} \,,\quad \mathcal{P}_{ij}\equiv \frac{1}{L^3}\sum_{\bq} \mathcal{P}_{\mu(i)\nu(j)}(\bq) e^{i\bq\cdot\bm{r}_{ij}}\,,
    \label{eq:SM-Delta-ij}
\end{equation}
where $\mu(i)$ and $\nu(j)$ are the sublattices corresponding to sites $i$ and $j$ respectively, and the sum is over wavevectors given by \cref{eq:SM-allowed-q}. There are two ways to deal with the singularity at $\bq=\bm{0}$: either remove $\bq=\bm{0}$ from the sum and replace $L^3$ with $L^3-1$, or shift every $\bq$ point slightly to avoid the zone center, e.g. by $\delta\bq \equiv (\bm{b}_1+\bm{b}_2+\bm{b}_3)/2L$. The above formula for $\Delta_{ij}$ works so long as $L$ is sufficiently large relative to the separation of sites $i$ and $j$. The Fourier transformed dipolar interaction matrix can then be written
\begin{equation}
    \mathcal{D}_{\mu\nu}(\bq) = c\mathcal{P}_{\mu\nu}(\bq)+ \sum_{n}\Delta_n A^{(n)}_{\mu\nu}(\bq),
    \label{eq:SM-Delta-mu-nu}
\end{equation}
where $\Delta_n$ is the value of $\Delta_{ij}$ for neighbor type $n$. We compute the values of $\Delta_n$ via \cref{eq:SM-Delta-ij}, truncating the sum in \cref{eq:SM-Delta-mu-nu} at $\abs{\bm{r}_{ij}} \approx 12 a_{\mathrm{nn}}$ as was done by \cite{SM-isakov2005}, using $L=60$. \cref{fig:SM-Delta-decay} shows that the calculated values of $\abs{\Delta}_{ij}$ decay as the inverse fifth power of distance, indicating convergence of \cref{eq:SM-Delta-ij}.


\newpage
\section{Additional Polarized Neutron Scattering Plots}
\label{SM-Additional-Results}
In this section, for the interested reader and to complement the main text, we provide additional polarized neutron scattering calculations for cases not studied therein.

\subsection{Temperature Dependence of the Flat NSF Intensity}
\label{SM-NSF-Temp-Dep}

\begin{figure*}[t]
	\centering
	\includegraphics[scale=0.9]{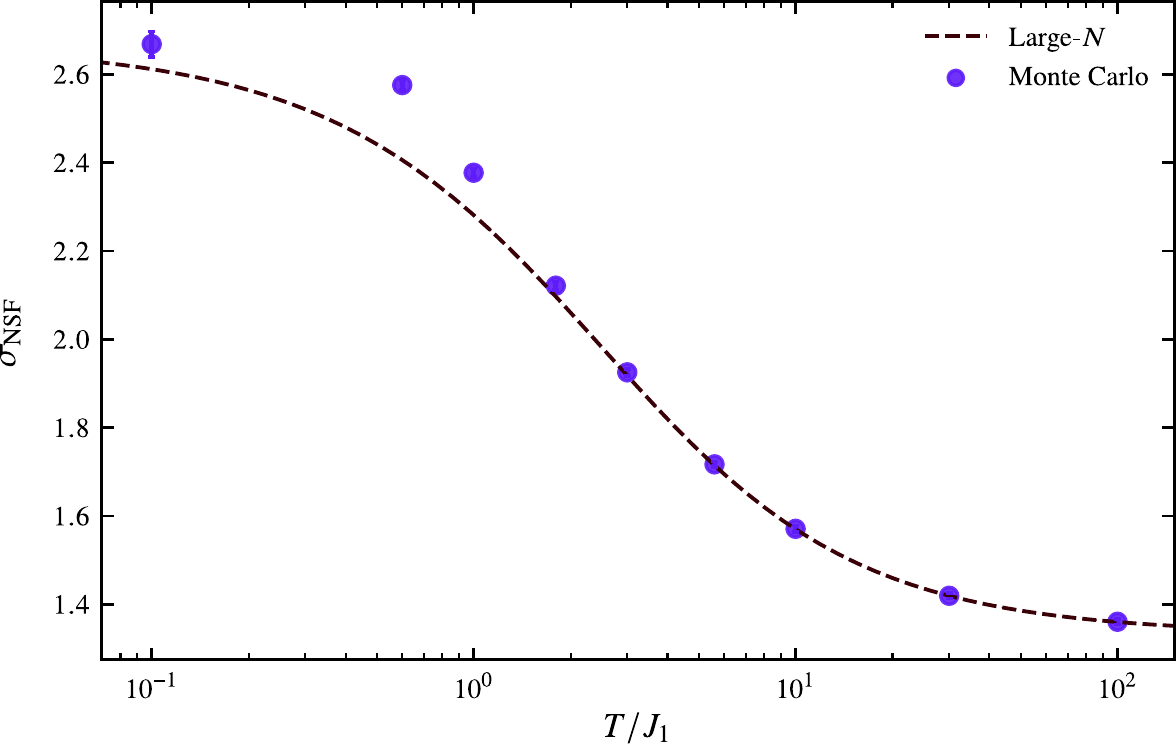}
	\caption{Evolution of the flat NSF values within temperature regime $0.1 < T/J_1 < 100$ for the NNSI model, obtained from large-$N$ approximation and Monte Carlo simulations.
	Error bars for the Monte Carlo results, if not visible, are smaller than the marker size.}
	\label{fig:nsf_temps}
\end{figure*}

It was noted in the main text that while the NSF is $\bq$-independent in the $(hhl)$ plane, its value is not the same at all temperatures $T$.
In \cref{fig:nsf_temps}, we show how the flat $\NSF$ value evolves with $T/J_1$ for the NNSI model, as predicted by large-$N$ ($=4/3\lambda$) and in our Monte Carlo simulations. The Monte Carlo calculation uses the definition of $\NSF$ in the main text Eq. (3), and could be quantitatively compared with calibrated neutron scattering experimental data, after dividing out the cross section prefactors discussed in \cref{SM-PNS}. $\NSF$ increases monotonically with decreasing $T$. The difference in temperature evolution between the large-$N$ and Monte Carlo is due to the fact that the Ising spin length constraint in Monte Carlo exponentially suppresses excitations at low energy, while the large-$N$/SCGA calculation softens the constraint and has a more gradual approach to the zero temperature limit.

\subsection{Large-\textbf{\textit{N}} Matrix Plots for SF and NSF Scattering}
\label{SM-Matrix-Plots}

\begin{figure*}[thbp]
    \centering
    \begin{overpic}[height=0.42\textheight]{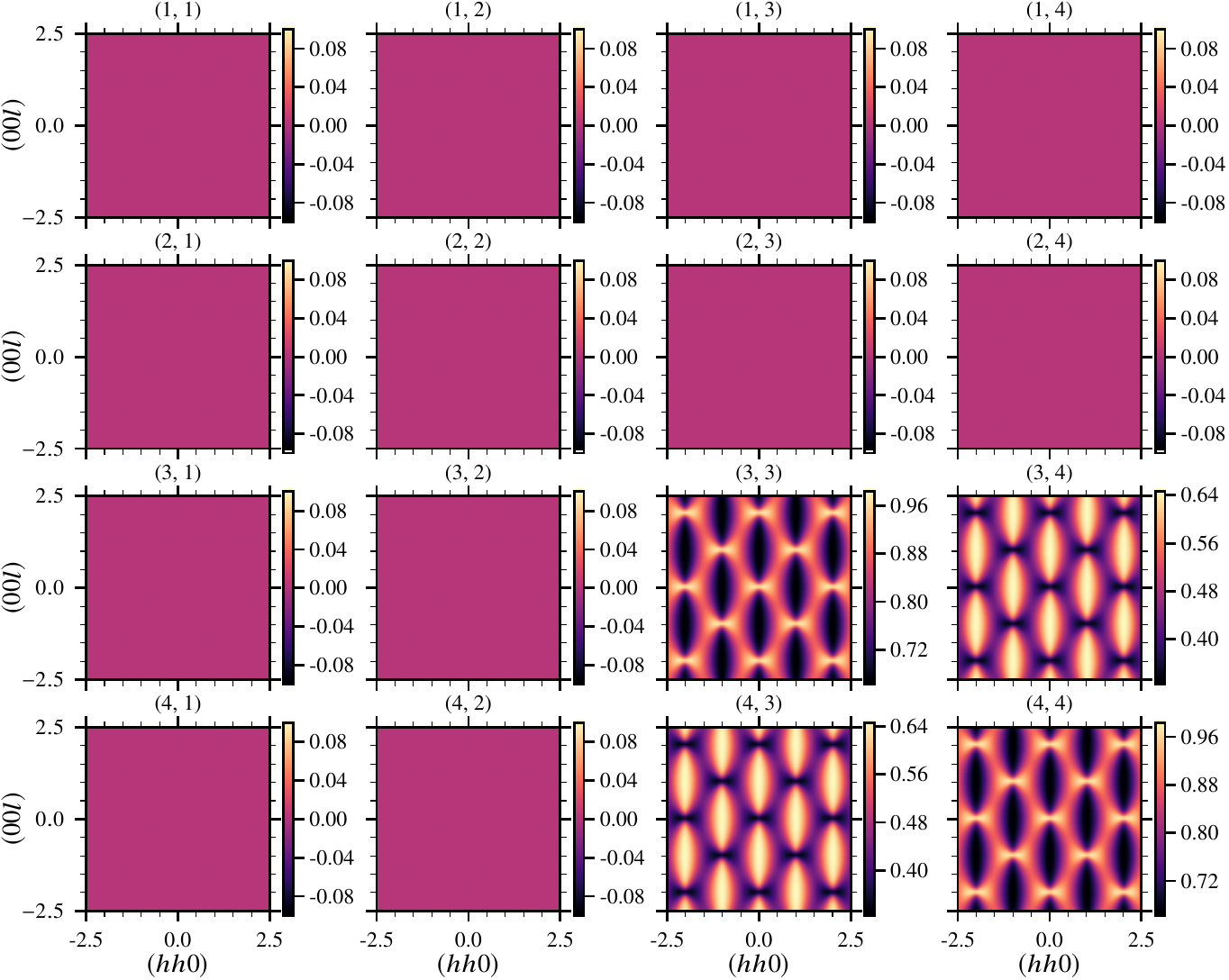}
    \put(-1.0,78.5){(a)}
    \end{overpic}\\
    \vspace{0.7cm}
    \begin{overpic}[height=0.42\textheight]{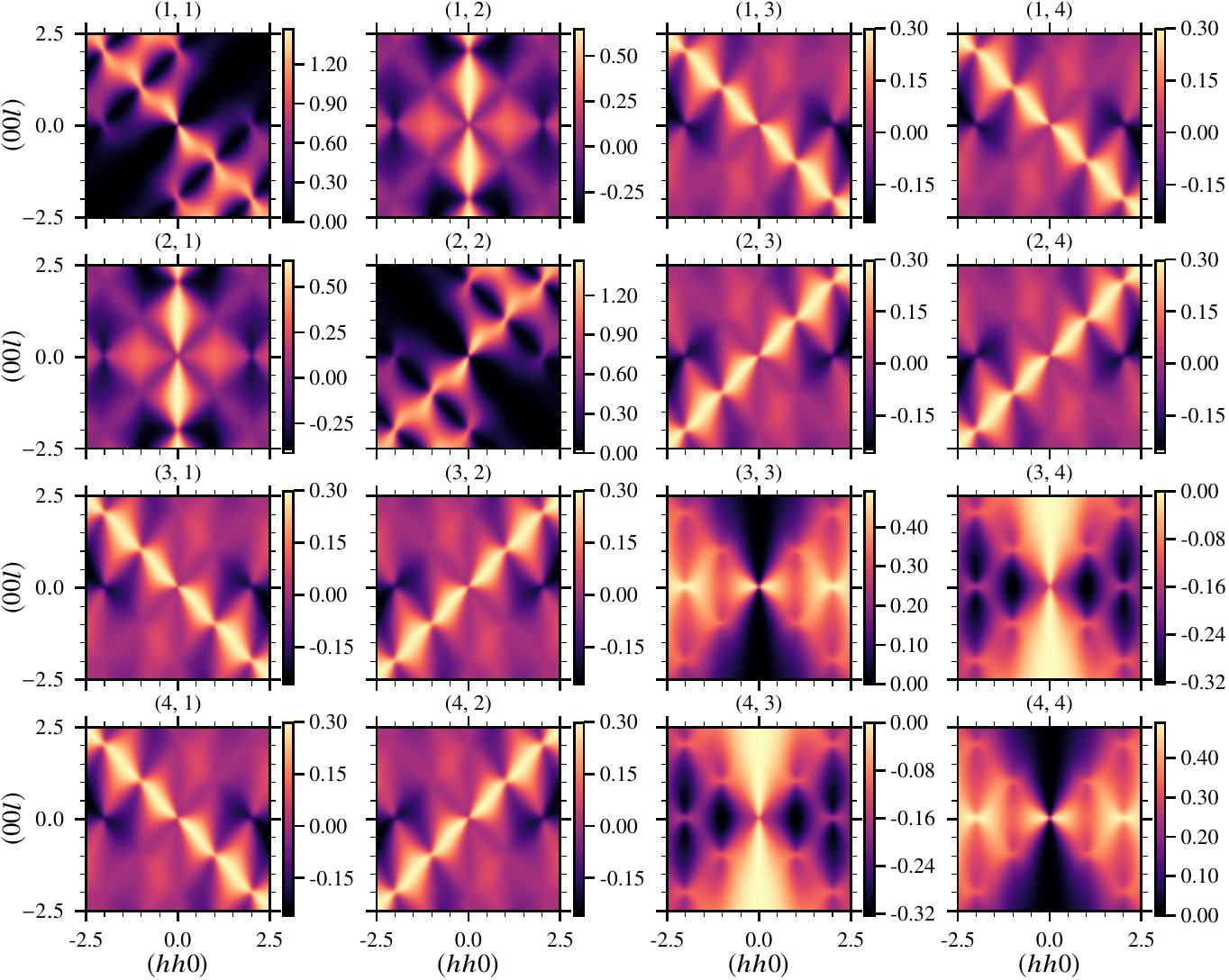}%
    \put(-1.0 ,78.5){(b)}
    \put(105,125){\large [NSF]}
    \put(105.25,40){\large [SF]}
    \end{overpic}
    \caption{Visualization of the full $4\times 4$ matrices of (a)~\cref{eq:G-NSF} (NSF) and (b)~\cref{eq:G-SF} (SF) cross sections at $T/J_{1}={0.1}$ for the NNSI. Twelve of the sixteen contributions to $\NSF$ vanish identically, while the remaining four are $\bq$-dependent, but exactly cancel when summed.}
    \label{fig:4x4gridplots}
\end{figure*}

We begin with \cref{fig:4x4gridplots} which shows the $4\times 4$ projected correlation functions for the NSF and SF cross sections in the $(hhl)$ plane, 
\begin{align}
\mathcal{G}^{\mathrm{NSF}}_{\mu\nu}(\bq) &\equiv (\uvec{z}_{\mathrm{sc}}\cdot\uvec{z}_\mu)\mathcal{G}_{\mu\nu}(\bq)(\uvec{z}_{\mathrm{sc}}\cdot\uvec{z}_\nu)\,, 
\label{eq:G-NSF}\\ 
\mathcal{G}^{\mathrm{SF}}_{\mu\nu}(\bq)  &\equiv (\uvec{y}_{\mathrm{sc}}\cdot\uvec{z}_\mu)\mathcal{G}_{\mu\nu}(\bq)(\uvec{y}_{\mathrm{sc}}\cdot\uvec{z}_\nu)\,,
\label{eq:G-SF}
\end{align}
with no implied summation. The total NSF and SF scattering functions, $\NSF(\bq)$ and $\SF(\bq)$ respectively, are given by summing all 16 elements of the corresponding matrix \cref{eq:G-NSF} or \cref{eq:G-SF}.
From \cref{fig:4x4gridplots}(a), and as discussed in the main text, it is evident that 12 elements of $\mathcal{G}^{\mathrm{NSF}}_{\mu\nu}(\bq)$ vanish by virtue of the NSF projection factors in  \cref{eq:nsf_projection_factors-hhl}. The four non-zero elements are equal to $\sqrt{2/3}$ times $\mathcal{G}_{33}$, $\mathcal{G}_{44}$, $-\mathcal{G}_{34}$, and $-\mathcal{G}_{43}$, and their sum is a constant. 
We have also provided a visualization of the SF matrix $\mathcal{G}^{\mathrm{SF}}_{\mu\nu}(\bq)$ in \cref{fig:4x4gridplots}(b), where the most striking difference from the NSF is that all 16 elements are non-zero and $\bq$-dependent.

\newpage
\subsection{NSF Scattering in (\textit{h0l}) Plane}
\label{SM-NSF-h0l}

\begin{figure*}[t]
    \centering
    \begin{overpic}[]{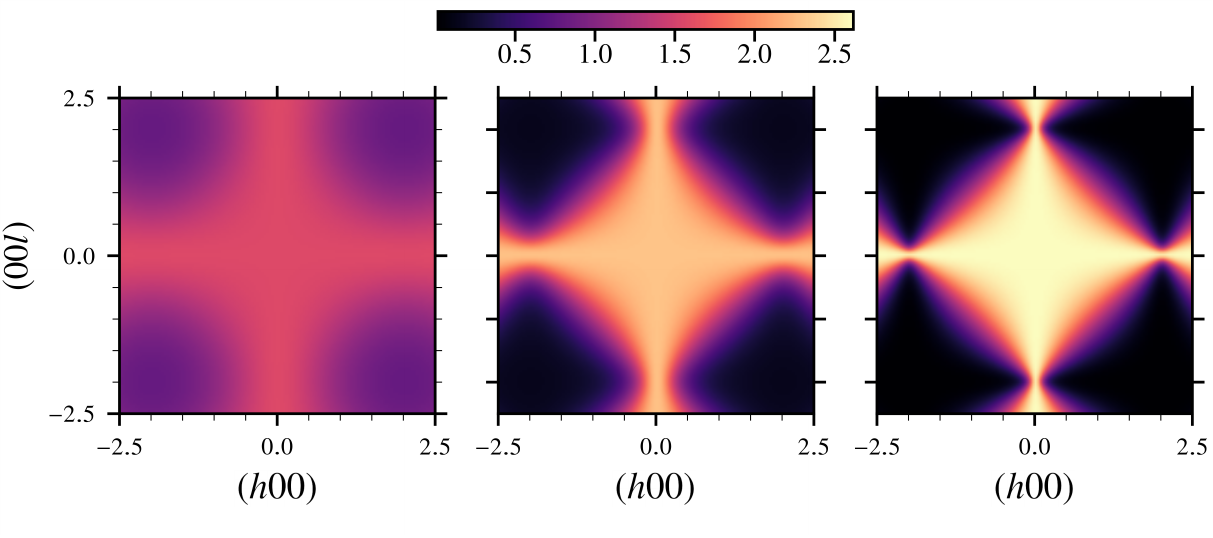}
    \put(11,33){\color{white}\footnotesize$T/J_1 = 10$}
    \put(42,33){\color{white}\footnotesize$T/J_1 = 1$}
    \put(73,33){\color{white}\footnotesize$T/J_1 = 0.1$}
    \end{overpic}
    \caption{NSF cross section for the $(h0l)$ scattering plane (with the corresponding neutron polarization direction $\uvec{z}_{\mathrm{s}}=[010]$) develops structure at all temperatures studied, including pinch points at $\{200\}$.}
    \label{fig:h0l_trio}
\end{figure*}

\begin{figure}[tbp]
    \centering
    \includegraphics[width=0.7\columnwidth]{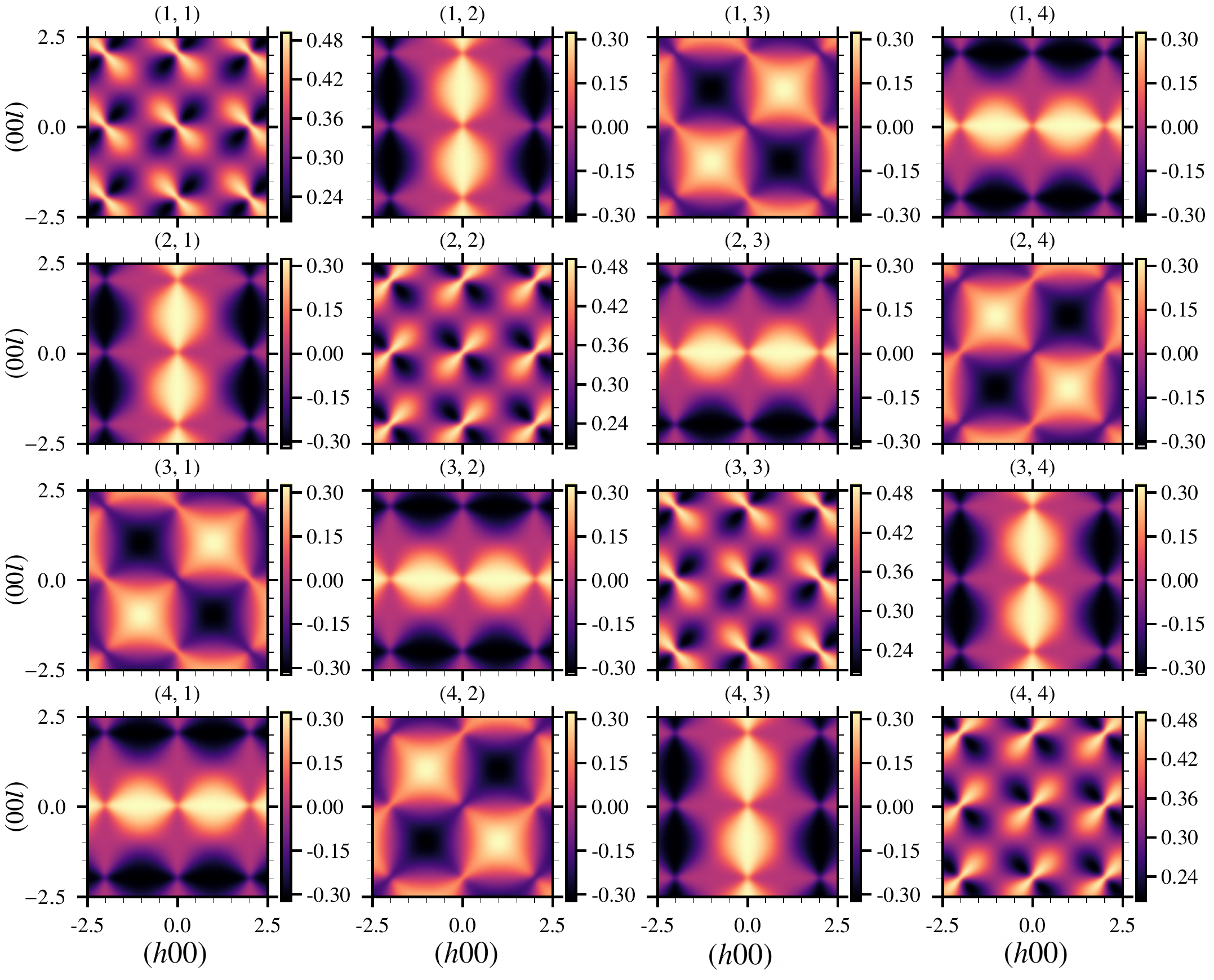}
    \caption{The elements of  $\mathcal{G}^{\mathrm{NSF}}_{\mu\nu}(\bq)$ in the $(h0l)$ scattering plane. All 16 are non-zero when performing polarized neutron scattering in this plane because of their non-vanishing NSF factors in \cref{eq:nsf_projection_factors-h0l}.}
    \label{fig:h0l_grid}
\end{figure}

The NSF cross sections in the $(h0l)$  plane with $\uvec{z}_{\mathrm{s}} \parallel [010]$ for NNSI are presented in \cref{fig:h0l_trio} at different temperatures $T/J_1 = 10, 1, 0.1$.
In contrast to the scattering in the $(hhl)$ plane (Fig.~1(f-h) of the main text), non-trivial structure develops in the NSF at all temperatures studied. For this scattering plane, the NSF projection factors are all non-zero,
\begin{equation}
    (\uvec{z}_{\mu} \cdot \uvec{z}_{\mathrm{s}}) = 
    \begin{cases}
        +1/\sqrt{3} &\quad \mu=1,3 \\
        -1/\sqrt{3} &\quad \mu=2,4 \\
    \end{cases}\,.
    \label{eq:nsf_projection_factors-h0l}
\end{equation}    
The respective sublattice contributions in $\mathcal{G}^{\mathrm{NSF}}_{\mu\nu}(\bq)$ that give rise to the dispersion in the NSF for this plane are shown in \cref{fig:h0l_grid}.

\subsection{Flat NSF with \texorpdfstring{$J_{1}<0$}{J1<0}}
\label{SM-NSF-FM}

\begin{figure*}[t]
    \centering
    \includegraphics[width=0.65\linewidth]{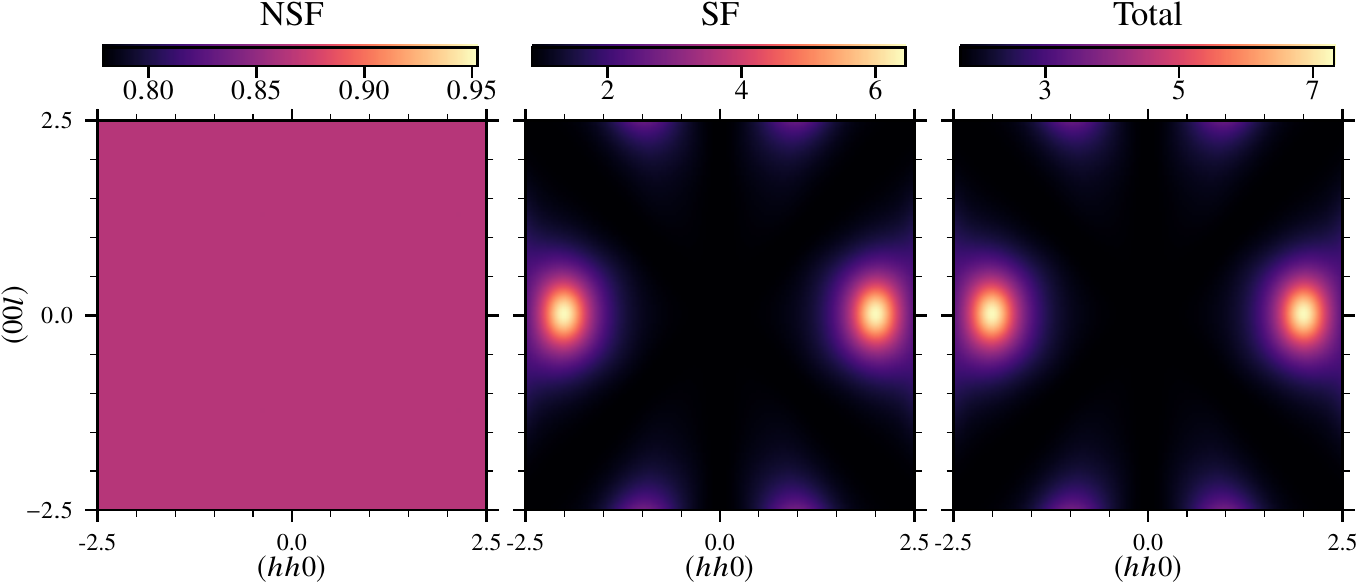}
    \caption{$\NSF(\bq)$, $\SF(\bq)$, and $\NSF(\bq)+\SF(\bq)$ in the $(hhl)$ scattering plane, calculated in the large-$N$ approximation for the NNSI model with $J_1<0$, at $|T/J_{1}|\sim 6$, slightly above the transition temperature to a long-ranged ordered state. The NSF remains completely $\bq$-independent, while Bragg peaks indicative of the phase transition appear in the SF scattering at $\{220\}$.
    }
    \label{fig:aiao_trio}
\end{figure*}

When $J_{1}<0$ in the nearest-neighbor Ising model, it becomes energetically favourable at low temperatures for the spins to take on an ordered all-in/all-out configuration~\cite{SM-Hertog2000}, instead of the two-in/two-out ``ice rules'' configurations.
We observe in \cref{fig:aiao_trio} that the flatness of the NSF channel persists even for $J_1<0$ for temperatures slightly above the transition temperature ($T/\abs{J_1} \approx 6$) to a long-ranged ordered state, while the strong scattering seen at $(220)$ in the total scattering appears \emph{exclusively} in the SF channel. Following the derivation in the main text (see also \cref{SM-Flat-NSF}), $\hat{\bm{\nsfvec}}$ is still an eigenvector of $V$, but now with eigenvalue $8J_1$ (instead of zero), so $\NSF(\bq)=(4/3)(\lambda + 8\beta J_1)^{-1}$ 
which is still flat, but with modified temperature dependence.

\subsection{ESI: MC Simulation and Large-\textit{N} Approximation Results}
\label{SM-ESI-MC}

In Fig.~1(c-h) of the main text, we showed the polarized neutron scattering results obtained from both Monte Carlo simulations and large-$N$ calculations for the NNSI model.
The analogous result for the extended spin ice (ESI) model with $J_{2} = J_{3a} \equiv J' = 0.1$ are presented in
\cref{fig:largeN_vs_MC_jprime01,fig:4x4gridplots_ESI01} below. The NSF remains flat at all temperatures, while the pinch points in the SF channel are broadened compared to the NNSI model at finite $T/J_1$ due to $J'$ lowering the energy of monopole excitations~\cite{SM-Rau2016}, therefore decreasing the crossover temperature at which the system enters the spin ice state.

\begin{figure*}[thbp]
	\centering
	\newcommand*{\figtwoLN}{\scriptsize{}\textcolor{black}{Large-$N$}}
	\newcommand*{\figtwoMC}{\scriptsize{}\textcolor{black}{Monte Carlo}}
	\begin{overpic}[]{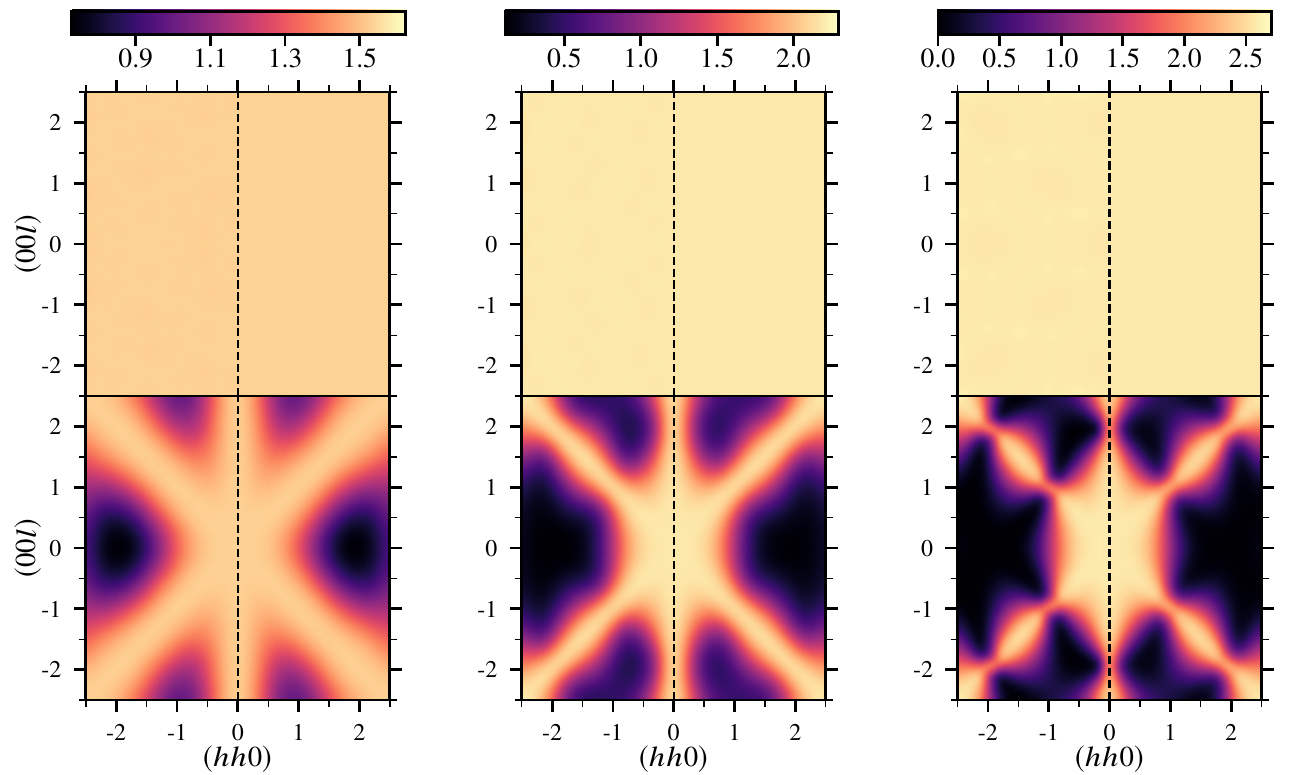}
	\put(7.5,30){\figtwoMC}
	\put(41,30){\figtwoMC}
	\put(74.5,30){\figtwoMC}
	\put(22.6,30){\figtwoLN}
	\put(56,30){\figtwoLN}
	\put(89.2,30){\figtwoLN}
    \put(7.5,49){\footnotesize $T/J_{1}=10$}
	\put(41,49){\footnotesize $T/J_{1}=1$}
	\put(74.5,49){\footnotesize $T/J_{1}=0.1$}
	\put(0,52){(a)}
	\put(34,52){(b)}
	\put(67.5,52){(c)}
	\put(0,27.5){(d)}
	\put(34,27.5){(e)}
	\put(67.5,27.5){(f)}
	\put(100,40){\large [NSF]}
	\put(100.25,17){\large [SF]}
	\end{overpic}
	\caption{%
	The polarized neutron scattering cross sections, NSF (top row) and SF (bottom row) (analogous to Fig.~1 from the main text) for the $J'=0.1$ ESI model conducted at (a,d)~$T/J_{1}=10$, (b,e)~$T/J_{1} = 1$ and (c,f)~$T/J_{1} = 0.1$.
	The left (right, resp.) half of each panel shows the Monte Carlo (large-$N$, resp.) results obtained in the $(hhl)$ plane.	$\NSF$ is seen to be featureless at all temperatures studied, with the flat value increasing monotonically as temperature is lowered. For fixed $T/J_1$, the SF scattering is slightly more diffuse than the NNSI model case.
	}
	\label{fig:largeN_vs_MC_jprime01}
\end{figure*}

\newpage

\begin{figure*}[h!]
    \centering
    \begin{overpic}[height=0.40\textheight]{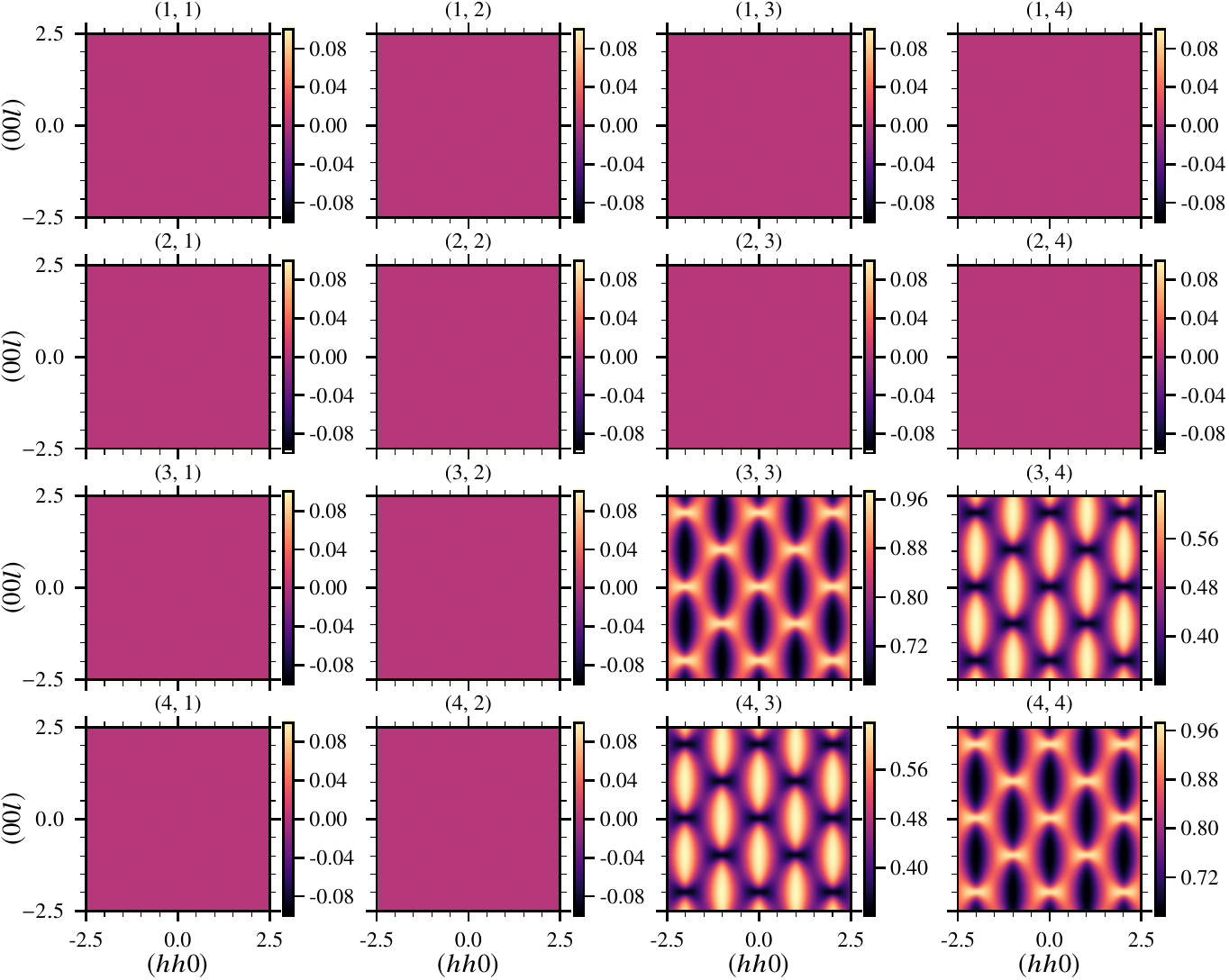}
    \put(-1,78.5){(a)}
    \end{overpic}\\
    \vspace{0.7cm}
    \begin{overpic}[height=0.40\textheight]{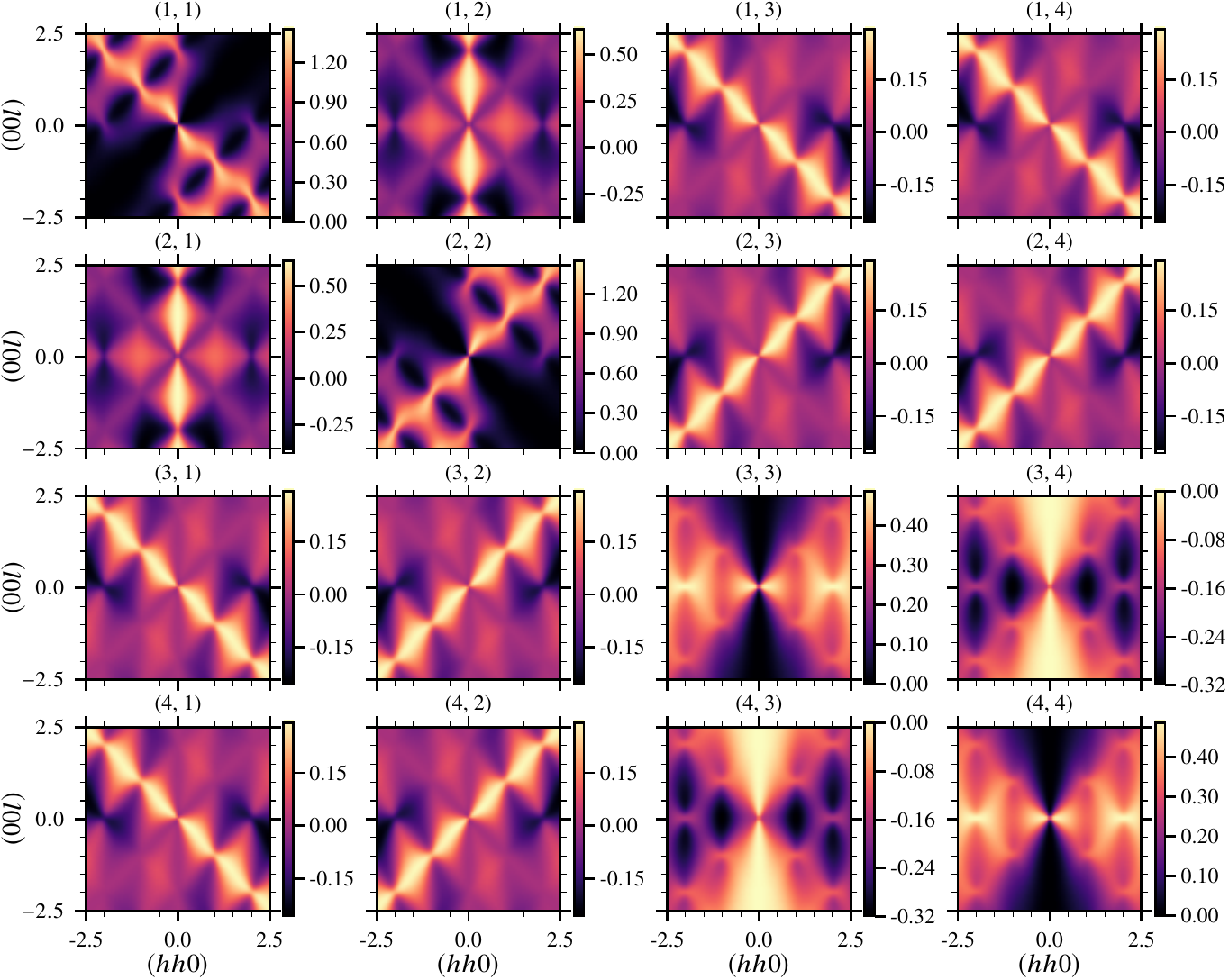}%
    \put(-1,78.5){(b)}
    	\put(105,125){\large [NSF]}
    	\put(105.25,40){\large [SF]}
    \end{overpic}
    \caption{Visualization of the full $4\times 4$ matrices of (a)~NSF and (b)~SF cross sections at $T/J_1={0.1}$ for the ESI model with $J'=0.1$.
    At such low $T/J_1$, the sublattice contributions are virtually identical to that of NNSI model in \cref{fig:4x4gridplots}.
    The NSF elements sum to yield a $\bq$-independent result.}
    \label{fig:4x4gridplots_ESI01}
\end{figure*}

\newpage
\bibliographystyle{apsrev4-1}

\begin{thebibliography}{62}
\bibitem [{foo()}]{footnote_FM}%
  \BibitemOpen
  \href@noop {} {}\bibinfo {note} {In \cite{SM} we show that the ferromagnetic
  $J_1<0$ model also exhibits a flat $(hhl)$ NSF in its paramagnetic
  phase.}\BibitemShut {Stop}%
\end{thebibliography}

\end{document}